\documentclass[modern]{aastex7}
\usepackage{CJKutf8}
\usepackage{multirow}

\newcommand{\K}{\ensuremath{\mathrm{K}}}

\newcommand{\cms}{\ensuremath{\mathrm{cm}\,\mathrm{s}^{-1}}}

\newcommand{\ergs}{\ensuremath{\mathrm{erg}\,\mathrm{s}^{-1}}}

\newcommand{\Msun}{\ensuremath{\mathrm{M}_\odot}}
\newcommand{\Ms}{{\ensuremath{{M}_{\odot} }}}

\newcommand{\cutt}[1]{\textcolor{blue}{}}
\newcommand{\code}[1]{{\texttt{#1}}}

\newcommand{\He}{{\ensuremath{^{4} \mathrm{He}}}}

\newcommand{\Hy}{{\ensuremath{^{1} \mathrm{H}}} }

\newcommand{\Figure}[1]{\textbf{Figure~\ref{fig:#1}}}

\newcommand{\Sectff}[1]{{\ref{sec:#1}}}
\newcommand{\Sect}[1]{{\S~\Sectff{#1}}}

\newcommand{\Eqref}[1]{{\ref{eq:#1}}}
\newcommand{\Eqff}[1]{{(\Eqref{#1})}}

\newcommand{\Equation}[1]{{Equation~\Eqff{#1}}}

\newcommand{\CASTRO}{{\texttt{CASTRO}}}
\newcommand{\FLASH}{{\texttt{FLASH}}}

\newcommand{\MESA}{{\texttt{MESA}}}

\newcommand{\CCSM}{\text{CSM}}

\newcommand{\mn}[1]{\textbf{\textcolor{black}{#1}}}
\newcommand{\mna}{\textbf{A}}
\newcommand{\mnb}{\textbf{B}}
\newcommand{\mnc}{\textbf{C}}
\newcommand{\mnd}{\textbf{D}}

\newcommand{\wy}[1]{{#1}}

\DeclareRobustCommand{\dl}{\bgroup\markoverwith{\textcolor{red}{\rule[.5ex]{2pt}{0.4pt}}}\ULon}

\begin{document}
\begin{CJK*}{UTF8}{bkai}
\title{Observable Signatures of Supernova Shock Breakout in Confined Circumstellar Medium}

\author[0009-0002-3816-4732]{Wun-Yi Chen（陳文翊）}
\affiliation{National Taiwan University, Graduate Institute of Astrophysics, Taipei, Taiwan, R.O.C.}
\affiliation{Academia Sinica, Institute of Astronomy and Astrophysics, Taipei 106319, Taiwan, R.O.C.}
\affiliation{Heidelberger Institut für Theoretische Studien, Schloss-Wolfsbrunnenweg 35, 69118 Heidelberg, Germany}
\email{wychen@asiaa.sinica.edu.tw}

\author[0000-0002-4848-5508]{Ke-Jung Chen}
\affiliation{Academia Sinica, Institute of Astronomy and Astrophysics, Taipei 106319, Taiwan, R.O.C.}
\affiliation{Heidelberger Institut für Theoretische Studien, Schloss-Wolfsbrunnenweg 35, 69118 Heidelberg, Germany}
\email{chenken1229@gmail.com}
\author[0000-0003-2611-7269]{Keiichi Maeda}
\affiliation{Department of Astronomy, Kyoto University, Kitashirakawa-Oiwake-cho, Sakyo-ku, Kyoto 606-8502, Japan}
\email{keiichi.maeda@kusastro.kyoto-u.ac.jp}
\author[0000-0002-4460-0097]{Friedrich K.\ R{\"o}pke}
\affiliation{Zentrum für Astronomie der Universität Heidelberg, Astronomisches Rechenzentrum, M{\"o}nchhofstr.\ 12--14, 69120 Heidelberg, Germany}
\affiliation{Heidelberger Institut für Theoretische Studien, Schloss-Wolfsbrunnenweg 35, 69118 Heidelberg, Germany}
\affiliation{Zentrum für Astronomie der Universität Heidelberg, Institut für Theoretische Astrophysik, Philosophenweg 12, 69120 Heidelberg, Germany}
\email{friedrich.roepke@h-its.org}
\begin{abstract}
Supernova shock breakout encodes rich information about stellar explosions, including the properties of the progenitor, the explosion mechanism, and the circumstellar environment. We present the first two-dimensional multigroup radiation-hydrodynamic simulations of shock breakout from Red Supergiant (RSG) progenitor stars embedded in confined circumstellar medium (\CCSM) produced shortly before core collapse. \wy{Our simulations reveal that radiation escaping ahead of the shock forms a radiative precursor that modifies the shock dynamics, drives multidimensional mixing within the radiating gas layers, and alters the emergent radiation from the \CCSM.}
\wy{For the CSM models explored here, the resulting bolometric light curves reach peak luminosities of ${\sim}1\text{--}3 \times 10^{44}\ \mathrm{erg\ s^{-1}}$ with durations of ${\sim}4\text{--}35$~hr, with these signatures controlled by the CSM mass and the confined radius.} Our multi-wavelength calculations further reveal a pronounced extreme-ultraviolet emission, where ionizing photons of energies ${\ge}54.4$~eV can sustain flash-ionized \ion{He}{2} and power strong emission from its recombination line of $\lambda 4686\ \text{\AA}$. Furthermore, we identify diagnostic signatures of shock breakout with dense \CCSM\ and assess their detectability with Einstein Probe and ULTRASAT. Our models and predictions provide promising diagnostics of shock breakout in \CCSM\ that may be detectable by current and forthcoming high-cadence ultraviolet and X-ray space telescopes.
\end{abstract}
\section{Introduction} \label{sec:intro}

The shock breakout (SBO) marks the first electromagnetic signal produced by the supernova (SN) explosion of a massive star. Its luminosity, duration, and spectral evolution encode valuable information about the progenitor radius, explosion properties, and immediate pre-explosion environment. Among massive-star progenitors, SBOs from red supergiants (RSGs) are particularly favorable for detection: their light curves (LCs) can reach peak luminosities of ${\approx}10^{44}$~\ergs\ and remain bright for several hours. By contrast, SBOs from blue supergiants (BSGs) and Wolf--Rayet (WR) stars typically last only tens of minutes, making them substantially more difficult to observe \citep[e.g.,][]{2008Schawinski, 2011Couch, 2015Gezari, 2017Waxman}. Direct detection of SN SBOs therefore requires wide-field X-ray monitoring with sub-hour cadence (${\lesssim}0.1$~hr), together with rapid ultraviolet (UV) follow-up observations over the subsequent day \citep{2010Gezari, 2022Bayless, 2023Hosseninzadeh, 2024Shrestha, 2024Shvartzvald, 2025Yuan}.

However, the observed LCs of SN SBOs often exhibit peak luminosities and durations that are inconsistent with the predictions of previous models \citep{1992Ensman, 2008Soderberg, 2011Chevalier, 2017Yaron, 2024Zimmerman}. Furthermore, a significant fraction of Type~II SNe display early-time flash-spectroscopy features characterized by transient, narrow emission lines, including $\mathrm{H}\alpha$, $\mathrm{H}\beta$, and high-ionization metal lines of C, N, and O, and, most notably, \ion{He}{2}~$\lambda 4686\ \text{\AA}$. These lines have characteristic velocities of ${\approx}100\ \mathrm{km\ s^{-1}}$ and likely originate from slowly expanding circumstellar material. Line emission can persist for ${\approx}4\text{--}5$~days before the spectra evolve into a nearly featureless continuum \citep{2021Bruch, 2023Bruch}. Their prevalence is unlikely to result from a selection bias toward unusually luminous events, because Type~II SNe exhibiting flash-ionization features are not systematically brighter at optical wavelengths than normal Type~II SNe \citep{2023Bruch}. Because the recombination timescales of He and highly ionized metal species are only a few minutes, much shorter than the observed spectral evolution timescale, the persistent line emission cannot be explained by a single, impulsive SBO flash. Instead, the lasting flash-spectroscopy features require a sustained source of ionizing radiation.

To explain the prolonged SBO LC and persistent narrow-line spectra, theoretical models commonly invoke a confined circumstellar medium (\CCSM) surrounding the SN progenitors \citep{2015Gezari, Hiramatsu_2023, 2024Dastidar}. When dense gas encompasses the exploding star, it effectively acts as an extended stellar envelope. Shock interaction with this material increases the photon-diffusion time and naturally broadens the SBO LC \citep{2017Lovergrove, 2020RSG_Alex, rsgwind, 2024Khatami,2026suzuki}. The radial extent of the \CCSM\ can be estimated from the duration of the flash-spectroscopy phase. For a characteristic SN shock velocity of $v_{\mathrm s}\approx10^{9}\ \cms$ and a flash-feature duration of ${\approx}5$~days, the inferred outer radius is $R_{\mathrm \CCSM}\approx v_{\mathrm s}\Delta t\approx4\times10^{14}$ cm. Reproducing the observed LCs typically requires an equivalent mass-loss rate of $\dot{M}\approx10^{-3}\text{--}10^{-1}\ \Ms\,\mathrm{yr}^{-1}$ for a wind velocity of $v_{\mathrm w}\approx10^{7}\ \cms$, sustained over the final $10\text{--}100$~yr before core collapse. Such \CCSM\ configurations imply density enhancements of approximately $2\text{--}5$ orders of magnitude relative to standard steady-state RSG winds \citep{2014Svirski, 2014Gal-Yam, 2022Margalit, 2024Jacobson-Gal, 2025Hu, 2026Sengupta, 2026Wasserman}. Several physical mechanisms have been proposed to generate such intense, short-lived mass loss immediately preceding core collapse, including late-stage interior instabilities and stellar pulsations that alter the envelope \citep{2025Bronner, 2026Laplace}, binary mass transfer or common-envelope interactions \citep{2025Matsuoka, 2026Tsai}, and wave-driven or pulsation-enhanced superwinds \citep{2026Sengupta}.

In addition to CSM, multi-dimensional (multi-D) geometric effects play a critical role in shaping early-time emission. Multi-D simulations reveal that radiation precursors (RPs) penetrating corrugated shock fronts (SFs) can smooth and extend the SBO signal by up to an order of magnitude compared to spherical 1D models \citep{suzuki2016, 2022Goldberg, 2024Chen}. Furthermore, RPs can radiatively accelerate slowly expanding CSM to $v_{\mathrm w} \approx 10^{7}\ \cms$ prior to shock arrival \citep{2024Zimmerman}. Consequently, multi-D geometry alleviates the extreme mass-loss rates otherwise required in 1D models to reproduce prolonged SBOs.

However, most multi-D SBO simulations still use single-group (grey) radiation transport, preventing direct predictions of spectral evolution and multi-band LCs \citep{suzuki2021,2022Goldberg}. Because both multi-D shock dynamics and CSM structure can broaden the breakout signal, single-band optical observations are intrinsically degenerate. Multi-wavelength coverage from X-rays to optical wavelengths is therefore essential to track the spectral energy distribution (SED) and color temperature evolution \citep{2008Gezari, Katz_2010, 2013Tolstov, 2018Foerster, 2022Margalit, 2022Goldberg, 2024Chen, 2024Zimmerman, 2026Wasserman}. Without dense \CCSM, such observations constrain envelope shock cooling and progenitor properties \citep{2011Chevalier, 2013Sapir}. With dense \CCSM, the resulting non-monotonic color evolution requires models that combine multi-D hydrodynamics with frequency-dependent radiation transport. Nevertheless, multigroup calculations remain largely limited to 1D and, therefore, cannot model the critical physical processes such as Rayleigh--Taylor mixing, directional photon leakage, and asymmetric CSM structure \citep{2017Lovergrove}.

We bridge this gap by conducting two-dimensional, 16-group radiation-hydrodynamics simulations of RSG SBOs embedded in \CCSM\ using the code \CASTRO\ \citep{2010Almgren, Almgren2020}. This framework enables us to simultaneously resolve multi-D hydrodynamic instabilities and compute multi-wavelength LCs, establishing a direct predictive bridge between multi-D models and early-time observational signatures.

The outline of this paper is organized as follows. We describe our progenitor models, \CCSM\ configurations, and numerical methods in \Sect{NM}. In \Sect{results}, we present our 2D simulation results and multi-wavelength LCs. We discuss the astrophysical implications in \Sect{dis}, and conclude our findings in \Sect{conclusions}.
\section{Numerical Methods}
\label{sec:NM}

\subsection{\CASTRO}
\label{subsec:castro}

We perform our two-dimensional (2D) multi-wavelength radiation-hydrodynamics (RHD) simulations using the \CASTRO\ code \citep{2010Almgren, Almgren2020}. \CASTRO\ is designed to model compressible astrophysical flows, employing a higher-order Godunov scheme to solve the hyperbolic hydrodynamic equations and the radiation frequency-space advection. The photon diffusion and radiation-matter source-sink terms are implicitly integrated using a first-order backward Euler method. To model radiation transport, the code adopts a comoving-frame multigroup flux-limited diffusion (MGFLD) solver \citep{2011Zhang, 2013Zhang}.

\wy{The RHD module in \CASTRO\ evolves the gas internal energy and the multigroup radiation energy densities independently, so the radiation field is not required to remain in local thermal equilibrium with the gas.} We close the RHD system of equations using the approximate flux limiter derived by \citet{1981Levermore}. For our frequency resolution of radiation, we discretize the radiation field into 16 logarithmically spaced groups spanning from $10^{14}$~Hz to $3\times10^{17}$~Hz. This frequency domain corresponds to photon wavelengths from ${\approx} 2\times10^4$~\text{\AA}\ (infrared, IR) down to ${\approx}10$~\text{\AA}, \wy{capturing the SED evolution of SBO across a broad wavelength range, from the infrared to soft X-rays.}

During SBO, the local gas temperature around the shock front (SF) exceeds $10^{5}$~K, fully ionizing \Hy\ and \He\ near the stellar surface and within the inner \CCSM. Consequently, the opacity in this regime is dominated by electron scattering. We therefore define the scattering opacity as $\kappa_{\mathrm s} = 0.4\, Y_{\mathrm e} \mathrm{\,cm^2\,g^{-1}}$, where $Y_{\mathrm e}$ is the electron fraction. To account for photon-electron energy exchange such as Compton scattering, we adopt an absorption opacity as $\kappa_{\mathrm a} = f \kappa_{\mathrm s}$, where $f = 10^{-4}$ is a fiducial value suggested by \citet{Ken2023ApJ}. This treatment follows the numerical methodologies suggested by \citet{2013Zhang} and \citet{2017Lovergrove}, and has been applied in our previous studies \citep{2024Chen,2026Chen}.

\subsection{Progenitor Stars and Explosion Setup}
\label{subsec:progenitor}

Our simulations begin with the RSG progenitor model adopted by \citet{2026Chen}. Using \MESA\ \citep{Paxton2011, Paxton2013, Paxton2015, Paxton2018, Paxton2019}, a $20\Msun$ zero-age main-sequence star with solar metallicity is evolved for approximately $8$~Myr. At the onset of core collapse, the model has a final mass of ${\approx}16\Msun$ and a stellar radius of $R_{\star}=7.69\times10^{13}$~cm. We then use \FLASH\ \citep{2000Fryxell} to initiate a 1D explosion by depositing thermal and kinetic energy within a narrow region near the iron-core/silicon-shell interface \citep{2020ono, 2024Ono}. An explosion energy of $1.09\times10^{51}$~erg yields a synthesized $^{56}\mathrm{Ni}$ mass of $4\times10^{-2}\,\Msun$. The calculation is continued until the outgoing shock enters the hydrogen-rich envelope.

At this stage, the 1D \FLASH\ profiles are mapped onto a 2D axisymmetric \CASTRO\ grid, and the remaining domain is initialized with the prescribed \CCSM\ structure described in Section~\ref{subsubsec:csm_geometry}. \wy{We define $t=0$ as the time when the 1D \FLASH\ profiles are mapped onto 2D \CASTRO\ grids to initialize the simulations.} We initialize the radiation field in local thermal equilibrium with a blackbody spectral distribution. To include convective motions already present in the progenitor, \wy{velocity perturbations} with amplitudes of approximately $15\%$ of the local fluid velocity are imposed throughout the stellar envelope \citep{2013Ken}. This choice is broadly consistent with earlier SN-mixing calculations that adopted velocity perturbations of ${\approx}30\%$ \citep{1998Nagataki} or density perturbations of ${\approx}25\%$ \citep{Mao_2015}. Because the small-scale structure of pre-explosion mass-loss flows remains poorly constrained, the \CCSM\ is initialized without perturbations.

The initial progenitor and explosion calculations retain spherical symmetry and therefore do not capture the multi-D dynamics of core collapse \citep[e.g.,][]{suzuki2016, 2022Nakamura, 2024Wang}. Nevertheless, after the profiles are mapped into \CASTRO, the 2D calculation naturally develops hydrodynamic instabilities, fluid mixing, and other multi-D structures generated during shock propagation and \CCSM\ interaction.

Our 2D simulations are performed in cylindrical coordinates $(r, z)$, assuming axisymmetry about the $z$-axis, and span a domain of $10^{15}\ \mathrm{cm} \times 10^{15}\ \mathrm{cm}$. The resulting spatial resolution is $\Delta r=\Delta z\approx2.44\times10^{11}$~cm, approximately an order of magnitude finer than the \wy{grid-cell size required for numerical convergence, as suggested by \citet{2024Chen}.} Reflective boundary conditions are applied along the symmetric axis at $r=0$ and $z=0$, whereas material is allowed to flow outward through the external boundaries. Each simulation covers approximately $60$--$100$~hr of physical evolution, encompassing shock propagation through the hydrogen envelope, breakout from the stellar surface, interaction with the \CCSM, and the subsequent rise and decline of the LC.
\subsubsection{CSM structure}
\label{subsubsec:csm_geometry}

Our CSM profiles consist of two distinct structural components: an ambient density from the steady RSG wind and an additional \CCSM\ structure on-top. For the ambient density, we adopt a constant mass-loss rate of $\dot{M} \approx 6 \times 10^{-5}\,\Msun\,\text{yr}^{-1}$ and a steady wind velocity of $v_{\mathrm w} \approx 3 \times 10^{6}\,\cms$ ($30\,\text{km\,s}^{-1}$) \citep{rsgwind}. The radial density profile $\rho(r)$ is modeled as a function of distance $r$ from the center of the progenitor star and parametrized by a density enhancement factor $\xi$:
\begin{equation}
\label{eq:csm_density}
\rho(r) = \frac{\xi \dot{M}}{4\pi\,r^2\,v_{\mathrm w}}.
\end{equation}

To construct realistic \CCSM\ profiles, we adopt observational constraints from \citet{2017Yaron, 2023Bruch,2024Zimmerman} and superimpose these structures onto the fiducial wind. While the \CCSM\ region responsible for flash spectroscopy may remain optically thin in certain SNe \citep{2023Bruch}, the dense CSM observed in events like SN~2023ixf can be optically thick \citep{2024Zimmerman}. We assume the intense mass-loss episode before the core collapse results in a dense, confined region extending directly from the stellar surface with an inverse-square density profile governed by \Equation{csm_density}. Specifically, we generate \CCSM\ profiles with density enhancement factors of $\xi = 5, 50,$ and $500$, terminating at outer shell radii of $R_{\mathrm \CCSM} \approx (1.6\text{--}2.0) \times 10^{14}$~cm. 

To model nonspherical CSM produced by interacting binaries, \citep[e.g.,][]{2026Tsai}, we adopt an equatorial torus with an elliptical cross section centered at $(r,z)=(1.5\times10^{14}\ \mathrm{cm},\,0\,\mathrm{cm})$, the ellipse of the torus has a 2:1 axis ratio, with its major axis aligned with the $r$ direction and spanning a full width of $10^{14}\ \mathrm{cm}$. The torus region has an enhanced density of $\xi=500$, embedded in a density profile of $\xi=5$ that extends to $2\times 10^{14}$ cm based on \Equation{csm_density}. Modeling this torus structure is motivated by early spectropolarimetric observations of SN~2023ixf, implying an aspherical distribution of CSM \citep{2026Nagao, 2026Vasylyev}. The model enables us to examine multi-D radiation transport, viewing-angle effects, and line-of-sight optical-depth variations in a low-mass asymmetric CSM. Physically, it represents an idealized circumbinary disk or equatorial outflow produced by late-stage binary interaction \citep{2026Tsai}.

\begin{figure}
    \centering
    \includegraphics[width=1.0\linewidth]{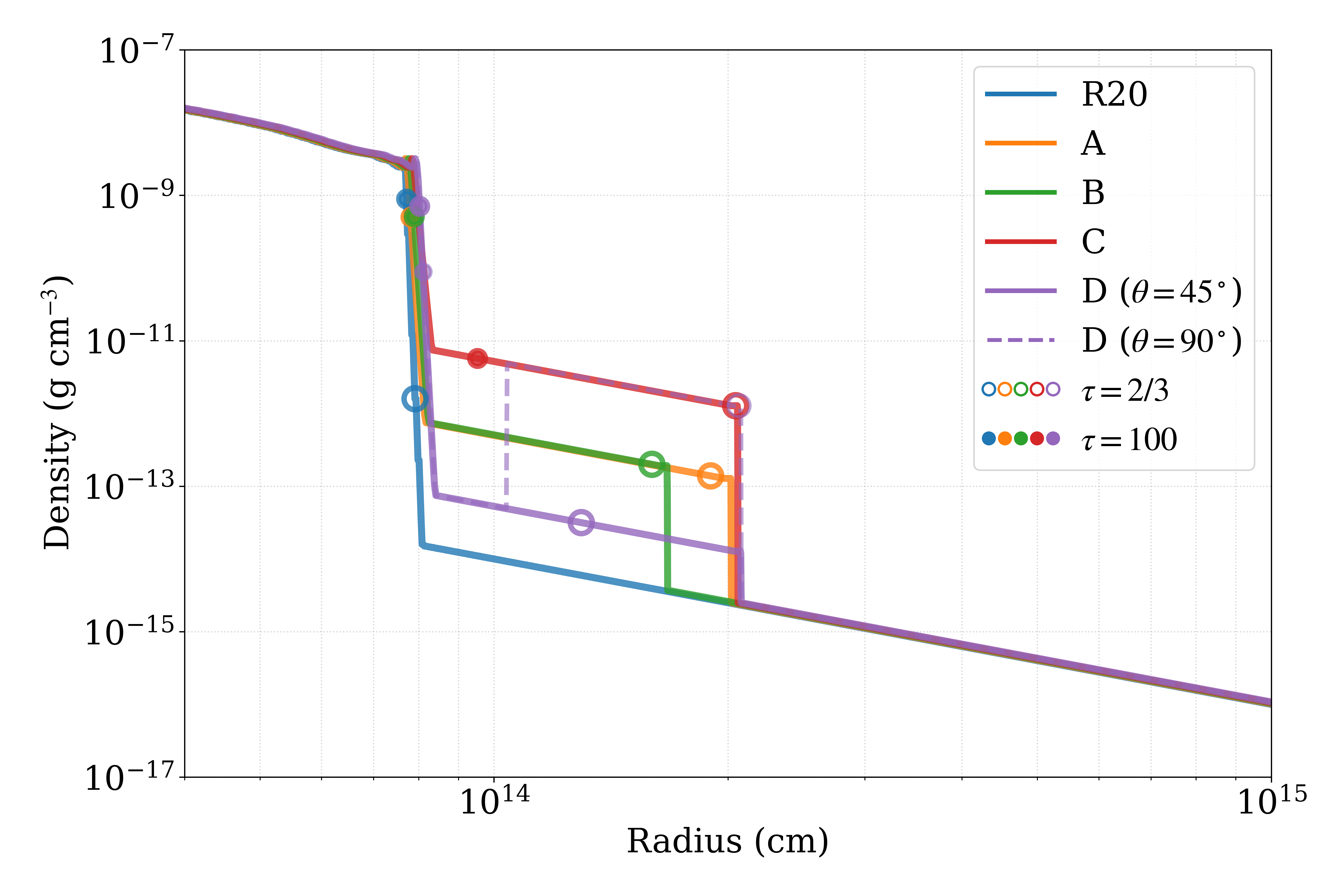}
    \caption{Initial CSM density profiles for all models. Open circles indicate the photospheric radius at $r(\tau=2/3)$, filled circles mark $r(\tau=100)$. Because model D has an aspherical CSM structure concentrated near the equatorial plane, profiles at viewing angles of $\theta=90^\circ$ (along the equator) and $45^\circ$ are shown. Model \mn{R20} represents \CCSM\ profile from the steady-wind derived from the original stellar-evolution model.
}
    \label{fig:1}
\end{figure}

The initial density profiles for all models are illustrated in \Figure{1} and their details are summarized in Table~\ref{tab:name}.

\begin{deluxetable*}{ccccc}
\tabletypesize{\scriptsize}
\tablewidth{0pt}
\tablecaption{Summary of \CCSM\ Model Parameters \label{tab:name}}
\tablehead{
\colhead{Model} &
\colhead{Structure} &
\colhead{$\xi$} &
\colhead{\CCSM\ Range ($10^{14}$~cm)} &
\colhead{$M_{\mathrm{CSM}}$ ($10^{-3}$~$\Msun$)}
}
\startdata
{\mna} & Sphere          & 50  & 0.8--2.0 & 4.0 \\
{\mnb} & Sphere          & 50  & 0.8--1.6 & 2.7 \\
{\mnc} & Sphere          & 500 & 0.8--2.0 & 40  \\
\multirow{2}{*}{\mnd} & Sphere & 5   & 0.8--2.0 & 0.4 \\
& Torus           & 500 & $r=1.0$--$2.0$; $z=0$--$0.25$ & 3.9 \\
\enddata
\tablecomments{
Model {\mnd} consists of a low-density spherical shell plus a dense equatorial
torus. \wy{Its total confined CSM mass is therefore
$M_{\mathrm{CSM}}\approx(0.4+3.9)\times10^{-3}\,\Msun
\approx 4.3\times10^{-3}\,\Msun$.
}}
\end{deluxetable*}
\subsection{Light Curve Calculations}
\label{subsec:lc_calc}
We derive observable color LCs from the 2D RHD simulations using the method of \citet{2024Chen}. At an extraction radius of $r_{\mathrm ext}\approx2.5\times10^{14}$~cm, we measure the magnitude of the frequency-dependent radiative flux, $|F_\nu|$, and calculate the monochromatic luminosity as $L_\nu=4\pi r_{\mathrm ext}^{2}|F_\nu|$. This radius remains well outside the photosphere at $r(\tau=2/3)$ throughout each simulation. Assuming cylindrical symmetry, we rotate the 2D data to construct a pseudo-3D emitting surface and apply light-travel-time corrections to every surface element. These corrections broaden and smooth the resulting LCs.

Because the breakout emission depends on the observer’s viewing angles relative to the asymmetric \CCSM\ structure, we define the viewing angle $\theta$ from the cylindrical symmetry axis $z$ and compare LCs between all models at $\theta=45^\circ$.

\begin{figure}
    \centering
    \includegraphics[width=1.0\textwidth]{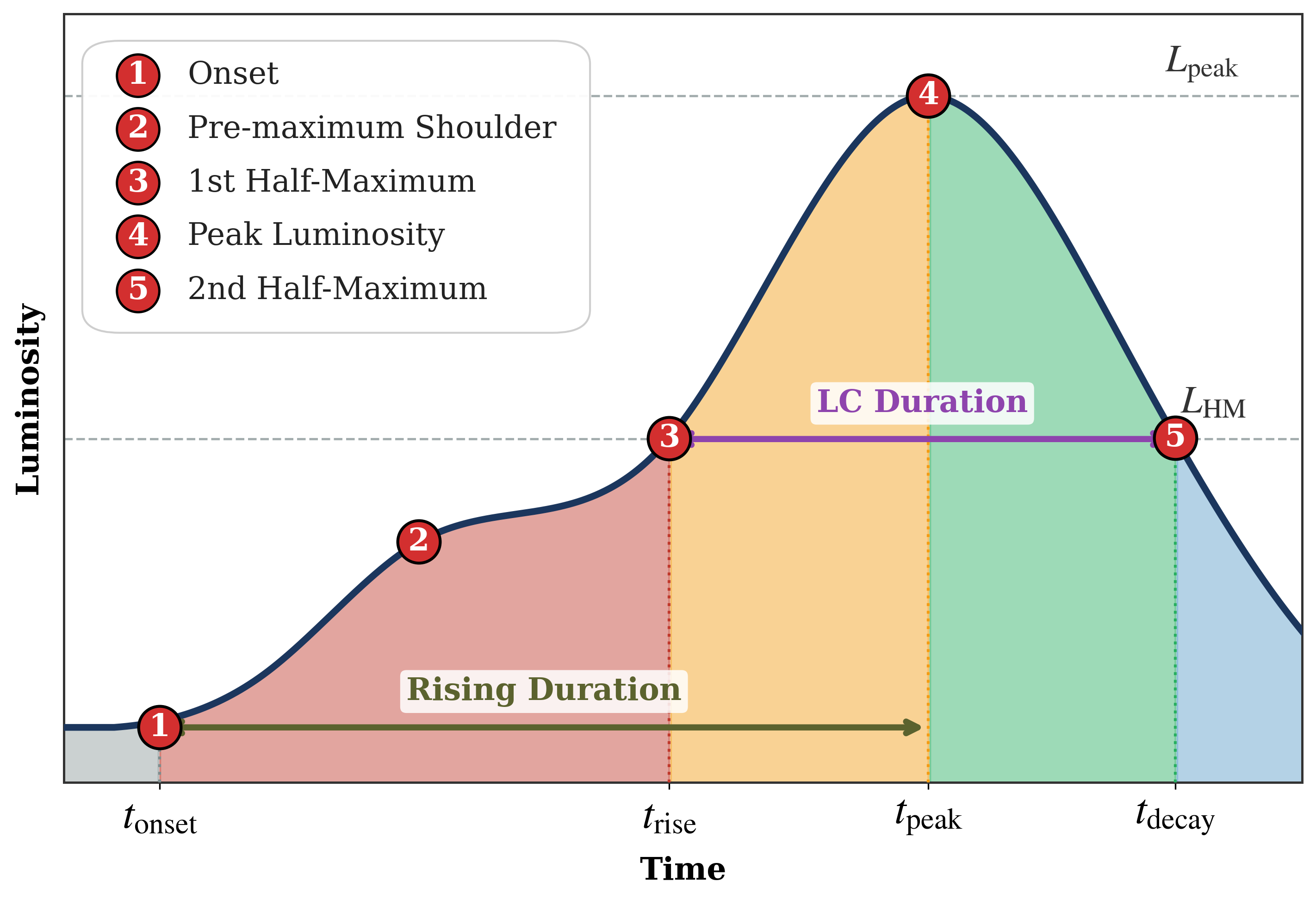}
    \caption{Schematic illustration of the characteristic LC features identified in our simulations. The cartoon marks luminosities of pre-maximum shoulder, half-maxima, and peak, indicating key timings for different evolutionary phases. We define the LC duration as $(t_{\mathrm {decay}}-t_{\mathrm {rise}})$.}
    \label{fig:2}
\end{figure}

A schematic illustration of the characteristic light-curve features is shown in \Figure{2}. Following \citet{2011Chevalier} and \citet{2022Goldberg}, we define the LC duration as the full width at half maximum, $t_{\mathrm {decay}}-t_{\mathrm {rise}}$, where $t_{\mathrm {rise}}$ and $t_{\mathrm {decay}}$ denote the rising and declining half-maximum crossings, respectively. If the LC does not reach $t_{\mathrm {decay}}$ before the simulation ends, we estimate its duration as $2(t_{\mathrm {peak}}-t_{\mathrm {rise}})$.

\subsection{Numerical Stability and Validation}
\label{subsec:verify}

To ensure numerical stability, we smooth the density discontinuities at the interfaces between the \CCSM\ structures and the ambient gas from the steady-state wind, using power-law buffer regions.

Each transition region has a characteristic width of ${\approx} 4\times10^{12}$~cm. Although this smoothing slightly perturbs hydrostatic equilibrium in the outer stellar atmosphere and may produce weak pre-breakout expansion or minor artifacts near the cylindrical symmetry axes, these effects are negligible compared with the energetics of shock \citep{2026Chen}.

\begin{figure}
    \centering
    \includegraphics[width=1.0\linewidth]{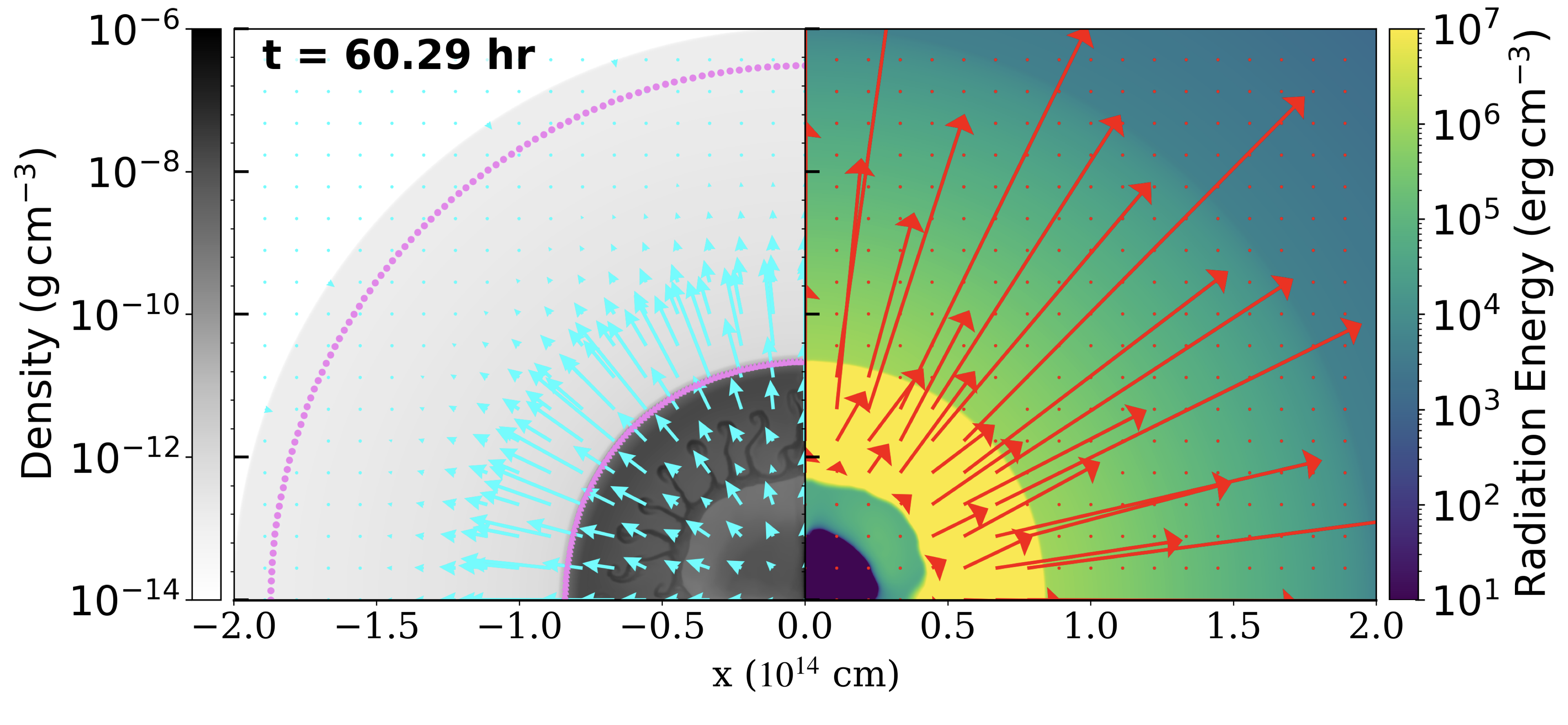}
    \caption{Snapshots of the gas and radiation energy-density distributions for model \mna\ at the moment when the shock breaks out of the stellar surface. Cyan and red vectors indicate the gas velocity and radiation flux, respectively. The inner and outer pink contours mark the optical-depth surfaces at $\tau=100$ and $\tau=2/3$. Rayleigh--Taylor instabilities are visible near the inner shocked envelope at $r\approx6\times10^{13}$~cm.}
    \label{fig:3}
\end{figure}

As shown in \Figure{3}, neither the velocity nor the radiative flux exhibits significant growth in the interfaces between the stellar envelope, \CCSM, and ambient gas before the SBO. The imposed density gradients therefore do not generate appreciable spurious radiation, alter the shock propagation, or affect the resulting emission.

\section{Results}\label{sec:results}
\subsection{Gas Dynamics and Radiation Precursors}
\label{subsec:gas_dynamics}

\begin{figure}
    \centering
    \includegraphics[width=1.0\textwidth]{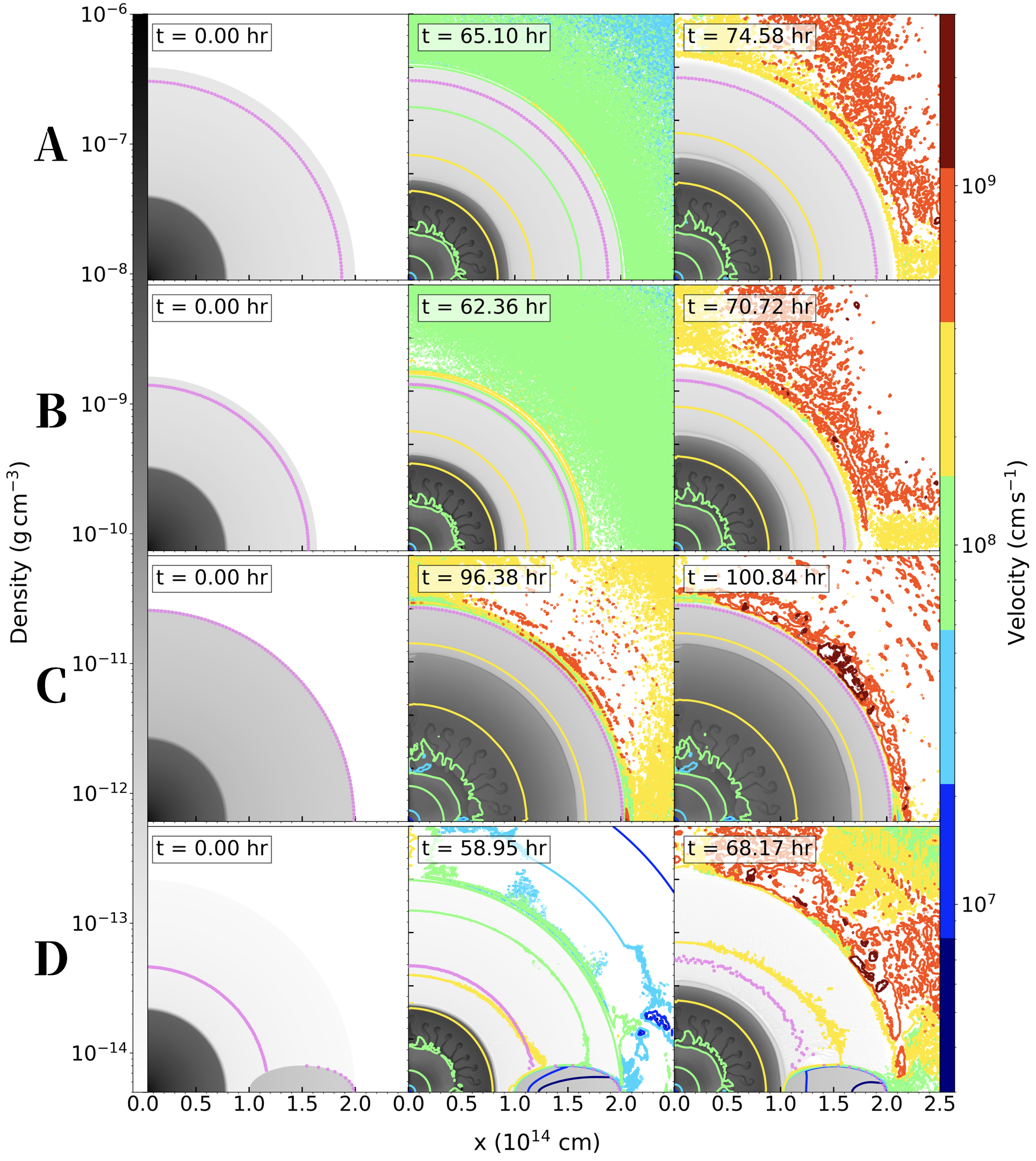}
    \caption{Density distributions overlaid with radial-velocity contours for all models. From left to right, the columns show the \wy{initial state}, peak luminosity, and post-peak phase. The pink contour marks the photosphere as seen by a distant observer. At peak luminosity, strong velocity variations develop beyond the confined CSM, showing the dynamical response of the outer gas to the radiation precursor. Relative to a bare RSG, the presence of a \CCSM\ extends the photospheric radius and prolongs the overall SBO emission.
}
    \label{fig:4}
\end{figure}

We first show the 2D density and velocity distributions of models \mna, \mnb, \mnc, and \mnd\ in \Figure{4}. In the left column, the SN shock remains deep within the stellar envelope and well below the photosphere. \wy{As the shock evolves, radiation diffuses ahead of the SF, forming a radiation precursor (RP) that heats and accelerates the upstream gas and \CCSM. Density and velocity inhomogeneities develop before SF arrival \citep{2008Schawinski, 2024Chen}. Meanwhile, the inner shocked envelope develops finger-like structures in regions susceptible to hydrodynamic Rayleigh--Taylor (RT) instability.} 

\begin{figure}
    \centering
    \includegraphics[width=1.0\linewidth]{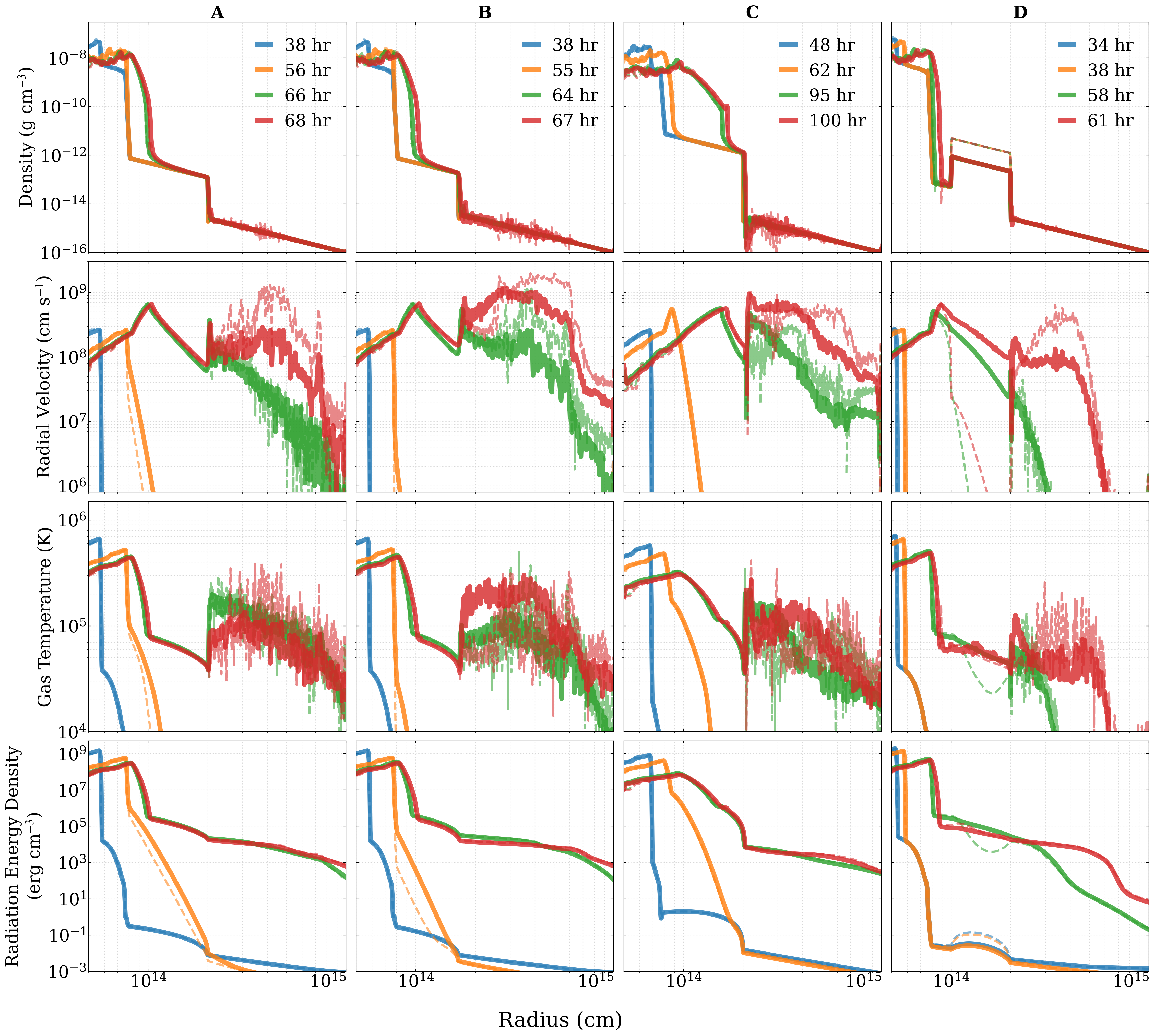}
    \caption{Evolution of the radial density, velocity, gas temperature, and radiation energy density profiles from pre-breakout to post-breakout phases. \wy{We extract these quantities along sampled radial directions through the 2D CASTRO domain. Solid curves show profiles averaged over the sampled radial directions, while dashed curves represent the radial profile near the equatorial plane.} As the radiation precursor heats and accelerates the CSM, the angle-dependent profiles increasingly depart from spherical symmetry, producing substantial spatial variations in velocity and temperature.}
    \label{fig:5}
\end{figure}

\wy{\Figure{5} presents the evolving radial profiles of gas density, velocity, temperature, and radiation energy density. Radiation propagating ahead of the SF modifies the shock dynamics by altering the velocity structure and thermodynamic properties of the upstream gas. As shock breakout proceeds, radiation propagates progressively outward beyond the confined CSM, heating the low-density gas to high temperatures. Meanwhile, pronounced angular variations develop in the velocity and temperature profiles, revealing the emergence of multidimensional mixing during shock breakout.}

Model \mnd\ exhibits the strongest angular dependence and largest fluctuations because of its torus \CCSM\ geometry. The dense structure absorbs and redirects outward-propagating radiation, driving shear instabilities along its boundaries and channeling the radiative flow around the equatorial region. Ablated material is subsequently accelerated outward, distorting the spherical photosphere.

\subsection{Light Curves of Shock Breakout}
\label{subsec:LCs}

\wy{In our 2D RHD calculations, shock breakout is an extended radiative process rather than an instantaneous event, even for models without an optically thick CSM \citep{2024Chen,2026Chen}.} 
\wy{Photons begin to diffuse ahead of the SN shock when the outward optical depth becomes comparable to $\tau\approx c/v_{\mathrm{s}}\approx 100$. The escaping radiation forms an RP that propagates toward the photosphere at $r(\tau=2/3)$. The breakout emission is sustained by the release of trapped radiation, with its duration governed by the evolving diffusion as the forward shock and RP propagate outward through a decreasing intervening optical depth.} Hydrodynamic mixing, the radial extent of the stellar envelope, and the \CCSM\ density distribution all strongly influence this timescale.

\begin{figure}
    \centering
    \includegraphics[width=\linewidth]{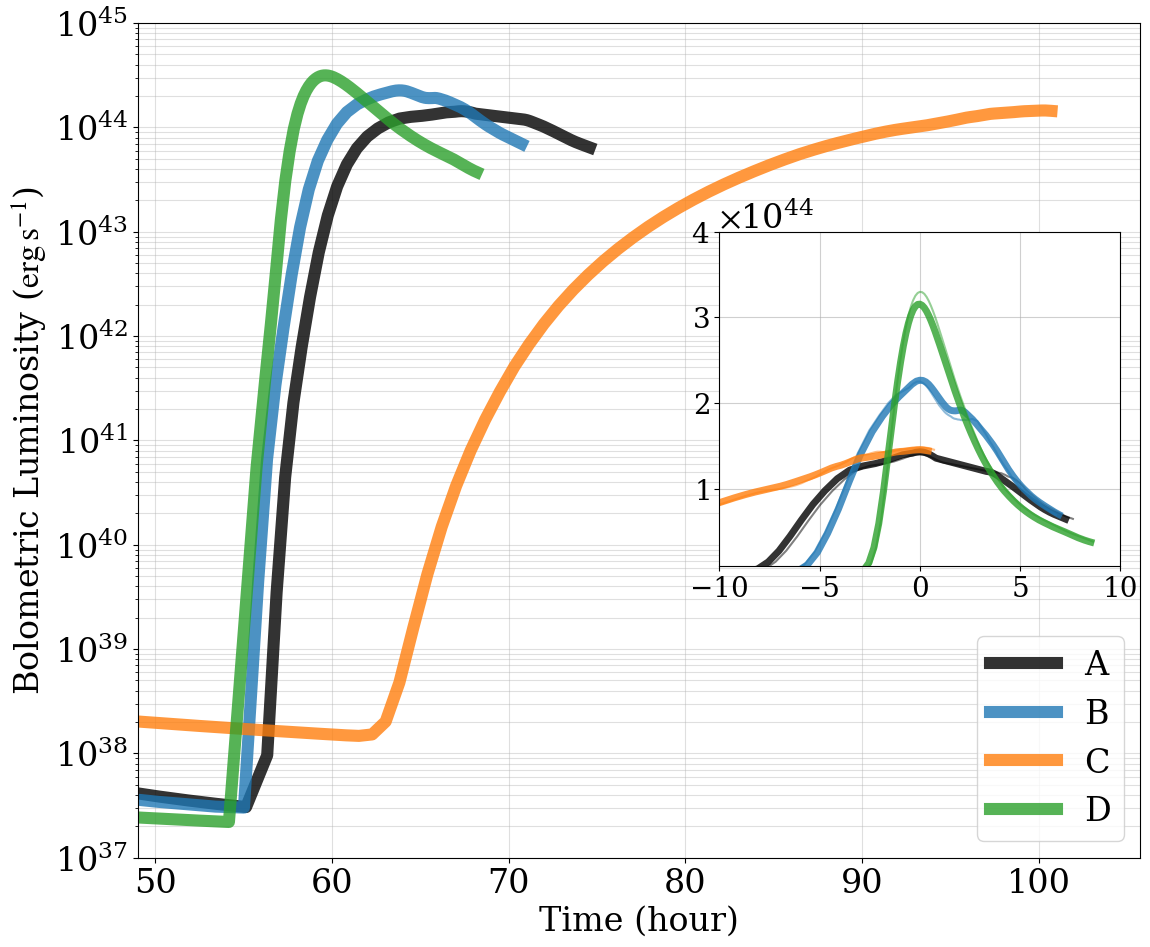}
    \caption{Bolometric LCs from the $\theta = 45^\circ$ viewing angle for all models. The rising durations ( $t_{\mathrm {peak}}-t_{\mathrm {onset}}$) are 12.4, 8.90, 40, and 5~hr for models \mna, \mnb, \mnc, and \mnd, respectively. \wy{The zoom-in subplot aligns all peaks at $t=0$, illustrating the peak structure and luminosities from $\theta = 90^\circ$ (thin lines) and $45^\circ$ (thick lines, \mnd\ $>$ \mnb\ $>$ \mna).}}
    \label{fig:6}
\end{figure}

The bolometric LCs for all models are shown in \Figure{6}. \wy{Among all models, the peak luminosities of LCs are ${\approx}(1$--$3)\times10^{44}$~erg~s$^{-1}$ with durations ${\approx}4$--$35$~hr.} The \CCSM\ delays and lengthens the onset of breakout. Despite the multi-D gas dynamics, the bolometric emission remains largely quasi-spherical and exhibits only modest viewing-angle dependence. Model \mnd\ is the clearest exception, showing an approximately $10\%$ difference in peak luminosity between $\theta=90^\circ$ and $\theta=45^\circ$ viewing angles. This asymmetry arises because the presence of the torus \CCSM\ shifts the photosphere to larger radii along the midplane than at higher latitudes.

\begin{figure}
    \centering
    \includegraphics[width=1.0\linewidth]{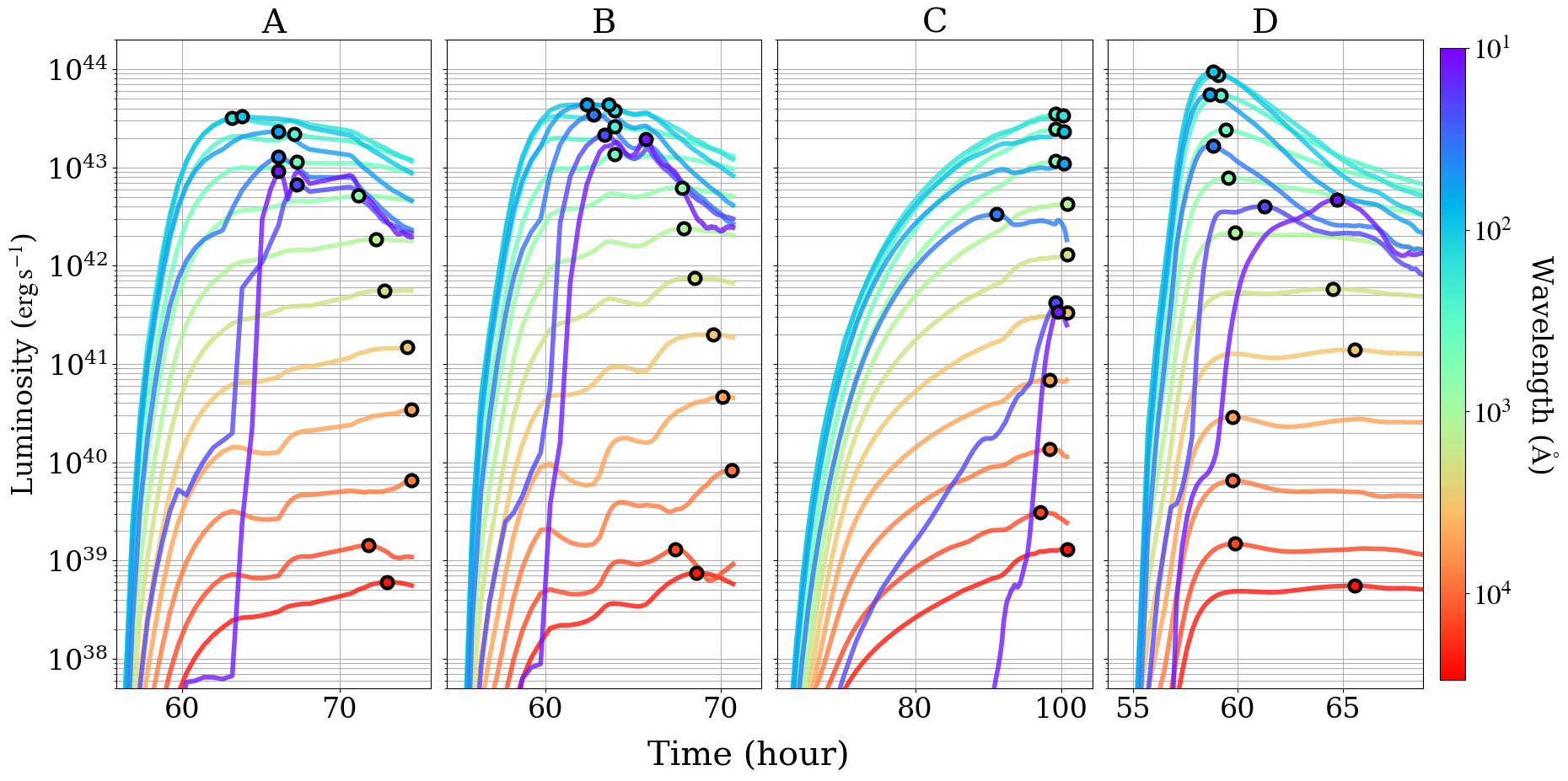}
    \caption{Multigroup LCs for all models, calculated at a viewing angle of $\theta=45^\circ$. Circles mark the peak luminosities. The breakout emission begins rising sharply approximately $5$--$15$~hr before maximum light and subsequently declines through gradual radiative cooling. In the less dense \CCSM\ models (\mna, \mnb, and \mnd), the soft X-ray emission exhibits a bumpy rise and peaks later than the UV emission. Increasing the \CCSM\ density generally broadens the LC peaks.
}
    \label{fig:7}
\end{figure}

\Figure{7} presents the multigroup color LCs. Their peak luminosities and LC durations are governed primarily by the radial extent and integrated optical depth of the \CCSM. Their peak luminosities and LC durations are summarized in Table~\ref{tab:2}.

The shortest-wavelength LCs generally exhibit three stages: a rapid rise, a primary peak, and a gradual decline. Most of the breakout energy emerges in the extreme-ultraviolet and soft X-ray bands, with the spectral flux peaking at wavelengths of ${\approx}50$--$200~\text{\AA}$. Approximately $5$~hr after the onset of the rise, a second peak emerges in the shortest-wavelength bands as deeper and hotter shock-heated material emerges through the \CCSM\ interface. In addition, broad shoulders appear in the soft X-ray LCs roughly $4$~hr before maximum luminosity, reflecting the progressive heating of the outer \CCSM\ by the RP. The simultaneous rise of the EUV and soft X-ray emission marks the main breakout epoch and coincides with its color evolution, after which radiative cooling becomes dominant.

\wy{Compared with the soft UV LCs, the X-ray LCs exhibit more pronounced variability on shorter timescales. This behavior may reflect the transient nature of high-energy emission, arising either from the initial RP or from subsequent shock interactions with the \CCSM, which produce energetic photons over relatively brief emission episodes.}

\begin{figure}
    \centering
    \includegraphics[width=1.0\linewidth]{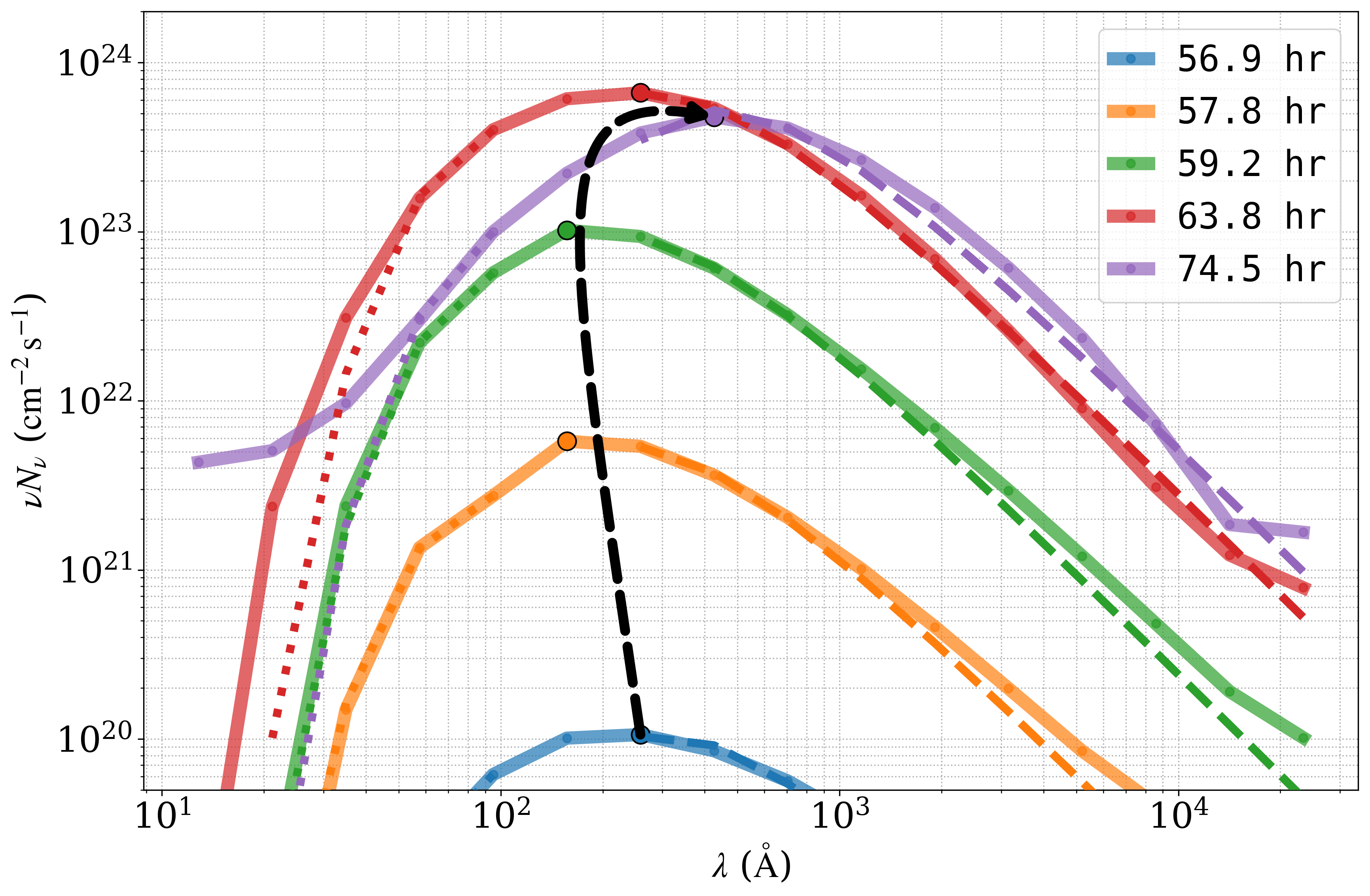}
    \caption{Evolution of the SED for model \mna\ during SBO. Colored curves represent distinct timing across the peak luminosity, circles mark peak flux wavelengths, and the black dashed curve indicates the trajectory of spectral peak evolution. Dotted and dashed lines show blackbody fits to the short-wavelength ($\lambda \lesssim 200\,\text{\AA}$) and long-wavelength regimes, respectively. Prior to peak luminosity, the spectral peak rises rapidly within ${\approx}5$~hr at $\lambda \approx 120\text{--}200\,\text{\AA}$. Post-peak, the flux peak shifts toward longer wavelengths ($\lambda \approx 200\text{--}350\,\text{\AA}$) while declining gradually in magnitude.}
    \label{fig:8}
\end{figure}

The temporal evolution of the SED for model \mna\ is shown in \Figure{8}. The spectrum hardens rapidly during the initial heating phase and reaches its highest characteristic energy near maximum luminosity. It subsequently softens more gradually as the shocked material undergoes radiative cooling.

Overall, our multi-D RHD simulations show that fluid instabilities, radiation-precursor dynamics, and variations in \CCSM\ structure jointly produce the broad diversity of SBO LCs.
\begin{deluxetable*}{cclcccccccc}
\tabletypesize{\scriptsize}
\tablecaption{
Characteristics of Multigroup LCs 
\label{tab:2}
}
\tablewidth{0pt}
\tablehead{
\multicolumn{3}{c}{} &
\multicolumn{2}{c}{\mna} &
\multicolumn{2}{c}{\mnb} &
\multicolumn{2}{c}{\mnc} &
\multicolumn{2}{c}{\mnd} \\
{\begin{tabular}[c]{@{}c@{}}Group \\ No.\end{tabular}} & 
\colhead{Band} &
\colhead{\begin{tabular}[c]{@{}c@{}}Center \\ (\AA)\end{tabular}} & 
\colhead{\begin{tabular}[c]{@{}c@{}}Peak \\ ($\mathrm erg\,s^{-1}$)\end{tabular}} & 
\colhead{\begin{tabular}[c]{@{}c@{}}Duration\\ (hr)\end{tabular}} &
\colhead{\begin{tabular}[c]{@{}c@{}}Peak \\ ($\mathrm erg\,s^{-1}$)\end{tabular}} & 
\colhead{\begin{tabular}[c]{@{}c@{}}Duration \\ (hr)\end{tabular}} &
\colhead{\begin{tabular}[c]{@{}c@{}}Peak \\ ($\mathrm erg\,s^{-1}$)\end{tabular}} & 
\colhead{\begin{tabular}[c]{@{}c@{}}Duration \tablenotemark{a} \\ (hr)\end{tabular}} &
\colhead{\begin{tabular}[c]{@{}c@{}}Peak \\ ($\mathrm erg\,s^{-1}$)\end{tabular}} & 
\colhead{\begin{tabular}[c]{@{}c@{}}Duration \\ (hr)\end{tabular}}
}
\startdata
1 & IR & 18176.81 & $6.03 \times 10^{38}$ & 30.45 & $7.49 \times 10^{38}$ & 23.62 & $1.31 \times 10^{39}$ & 180.62 & $5.54 \times 10^{38}$ & 19.71 \\
 2 & IR &11023.13 & $1.42 \times 10^{39}$ & 28.12 & $1.30 \times 10^{39}$ & 21.26 & $3.10 \times 10^{39}$ & 63.49 & $1.48 \times 10^{39}$ & 7.33 \\
 3 & IR/Optical& 6681.39 & $6.52 \times 10^{39}$ & 32.61 & $8.28 \times 10^{39}$ & 26.84 & $1.37 \times 10^{40}$ & 61.14 & $6.51 \times 10^{39}$ & 7.10 \\
 4 & Optical& 4051.51 & $3.43 \times 10^{40}$ & 32.61 & $4.68 \times 10^{40}$ & 25.75 & $6.92 \times 10^{40}$ & 59.49 & $2.89 \times 10^{40}$ & 7.10 \\
 5 & NUV& 2456.86 & $1.48 \times 10^{41}$ & 32.10 & $2.01 \times 10^{41}$ & 24.67 & $3.31 \times 10^{41}$ & 62.46 & $1.40 \times 10^{41}$ & 18.10 \\
 6 & FUV& 1489.80 & $5.62 \times 10^{41}$ & 29.15 & $7.53 \times 10^{41}$ & 22.56 & $1.30 \times 10^{42}$ & 59.14 & $5.81 \times 10^{41}$ & 16.01 \\
 7 & FUV/EUV&  903.43 & $1.87 \times 10^{42}$ & 28.13 & $2.39 \times 10^{42}$ & 21.31 & $4.27 \times 10^{42}$ & 57.48 & $2.16 \times 10^{42}$ & 6.67 \\
 8 & EUV&  547.81 & $5.15 \times 10^{42}$ & 25.89 & $6.25 \times 10^{42}$ & 21.07 & $1.17 \times 10^{43}$ & 52.74 & $7.86 \times 10^{42}$ & 6.00 \\
 9 & EUV&  331.94 & $1.14 \times 10^{43}$ & 18.08 & $1.36 \times 10^{43}$ & 14.30 & $2.48 \times 10^{43}$ & 52.74 & $2.40 \times 10^{43}$ & 5.77 \\
10 &  EUV& 201.19 & $2.18 \times 10^{43}$ & 18.66 & $2.64 \times 10^{43}$ & 14.30 & $3.50 \times 10^{43}$ & 54.41 & $5.39 \times 10^{43}$ & 5.32 \\
11 &  EUV/soft X-ray& 122.00 & $3.21 \times 10^{43}$ & 10.77 & $3.84 \times 10^{43}$ & 14.30 & $3.36 \times 10^{43}$ & 58.02 & $8.68 \times 10^{43}$ & 5.08 \\
12 &  soft X-ray&  73.96 & $3.32 \times 10^{43}$ & 12.02 & $4.35 \times 10^{43}$ & 13.65 & $2.33 \times 10^{43}$ & 59.97 & $9.36 \times 10^{43}$ & 5.23 \\
13 &  soft X-ray&  44.84 & $2.35 \times 10^{43}$ & 16.64 & $4.42 \times 10^{43}$ & 11.13 & $1.09 \times 10^{43}$ & 59.97 & $5.49 \times 10^{43}$ & 4.96 \\
14 &  soft X-ray&  27.18 & $1.30 \times 10^{43}$ & 15.71 & $3.43 \times 10^{43}$ & 10.08 & $3.35 \times 10^{42}$ & 41.47 & $1.66 \times 10^{43}$ & 4.57 \\
15 &  soft X-ray&  16.48 & $6.76 \times 10^{42}$ & 8.24 & $2.16 \times 10^{43}$ & 6.24 & $4.20 \times 10^{41}$ & 32.33 & $4.03 \times 10^{42}$ & 6.87 \\
16 &  soft X-ray&  12.83 & $9.24 \times 10^{42}$ & 3.35 & $1.95 \times 10^{43}$ & 9.82 & $3.40 \times 10^{41}$ & 7.15 & $4.74 \times 10^{42}$ & 11.10 \\
\tableline
\multicolumn{3}{c}{\bf Bolometric LC} & $\mathbf{1.43 \times 10^{44}}$ & \bf 11.8 & $\mathbf{2.27 \times 10^{44}}$ & \bf 8.06 & $\mathbf{1.45 \times 10^{44}}$ & \bf 35.4 & $\mathbf{3.15 \times 10^{44}}$ & \bf 4.08 \\
\enddata
\tablenotetext{a}{The LC for this model does not decline below half of its peak luminosity before the simulation ends; we estimate its duration as $2\times(t_{\mathrm {peak}}-t_{\mathrm {rise}})$.}
\tablecomments{
Peak luminosities and LC durations measured at a viewing angle of $\theta=45^\circ$.}
\end{deluxetable*}
\section{Discussion}
\label{sec:dis}

\subsection{Confined CSM Effects on Shock Breakout Light Curves}
\label{subsec:dis_csm}

A confined CSM systematically broadens and prolongs SBO LCs relative to standard steady-wind models \citep{2026Chen}. This effect arises from the additional optical depth of the dense shell above the stellar surface.

\wy{Based on our models, we find an empirical relation between the time to peak and the \CCSM\ mass:}
\begin{equation}
\label{eq:2}
t_{\mathrm {peak}} \approx 57 + 0.83
\left(\frac{M_{\mathrm \CCSM}}{10^{-3}\,\Msun}\right)
\quad \mathrm{hr},
\end{equation}
where $M_{\mathrm \CCSM}$ is the \CCSM\ shell mass. The constant of $57$~hr represents the characteristic breakout timescale of our $20\,\Msun$ RSG model surrounded by an ambient gas density from a standard stellar wind. For the asymmetric \CCSM\ model \mnd, this relation approximately holds for \wy{$\xi=5$} because the early rise is governed primarily by the ambient steady-wind component rather than the torus.

This scaling relation can be understood from the photon-diffusion for the shock in \CCSM,
\begin{equation}
\wy{t_{\mathrm {diff}}\sim\frac{\tau R_{\mathrm \CCSM}}{c}
\sim\frac{\kappa M_{\mathrm \CCSM}}{4\pi cR_{\mathrm \CCSM}}
\propto  \frac{M_{\mathrm \CCSM}}{R_{\mathrm \CCSM}},}
\end{equation}
where $R_{\mathrm \CCSM}$ is the radial extent of the \CCSM, $\kappa$ is the gas opacity, and $c$ is the speed of light. The $R_{\mathrm \CCSM}$ is determined primarily by the wind velocity and the duration of the enhanced mass loss, whereas $M_{\mathrm \CCSM}$ depends on the rate and duration of mass loss. \wy{A larger \CCSM\ mass increases the optical depth and prolongs the escape of breakout radiation.}

\begin{figure}
    \centering
    \includegraphics[width=1.0\linewidth]{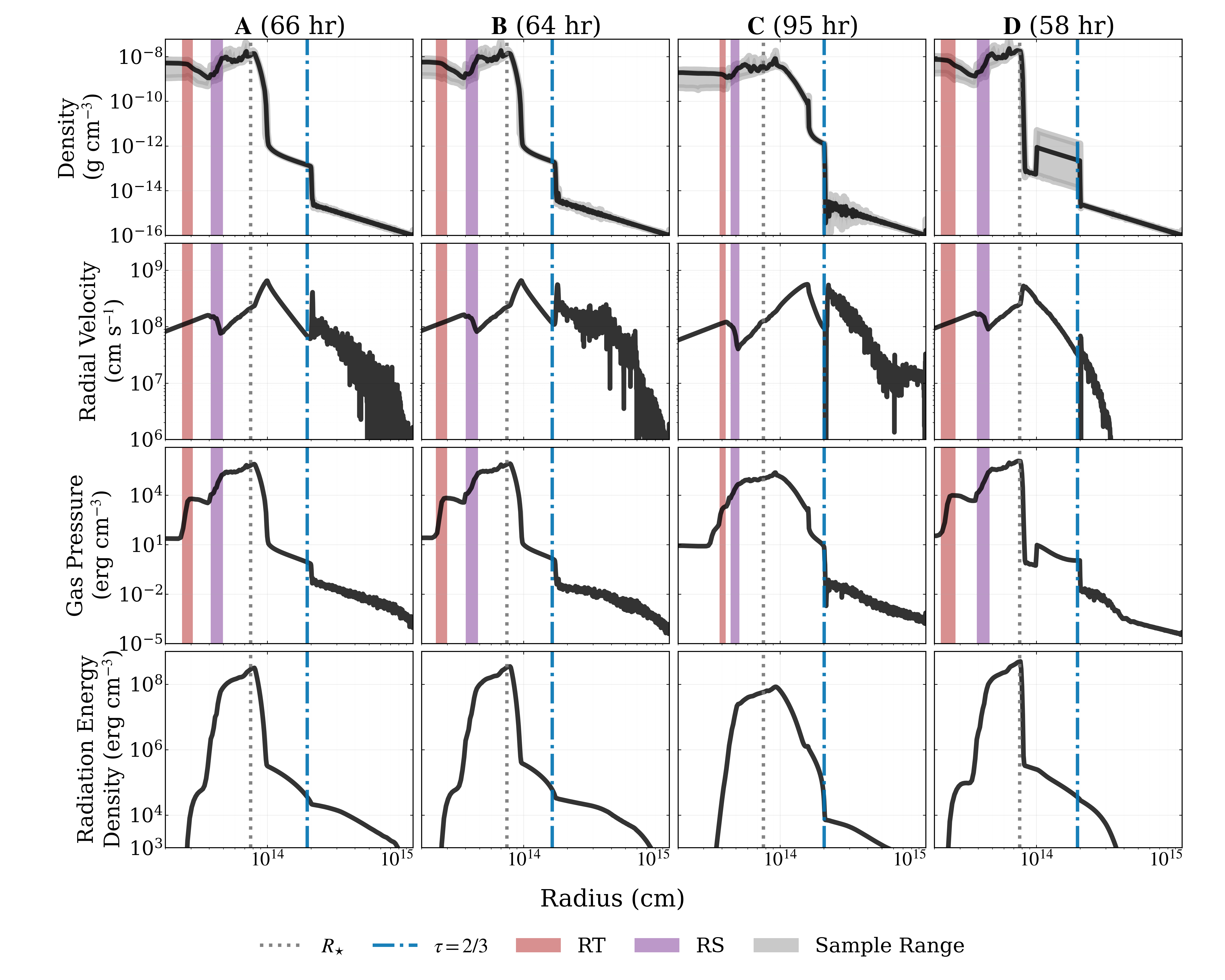}
    \caption{\wy{Angular-averaged radial profiles of physical quantities at the peak luminosity as the epochs marked by the green curves in \Figure{5}. From top to bottom, the panels show gas density, radial velocity, gas pressure, and radiation energy density. Black curves show angular-averaged value. Gray shading in the density panel covers sampled ranges of viewing angles. Black dotted and blue dash-dotted lines mark the initial stellar radius and the photosphere, respectively. Red shading indicates regions susceptible to hydrodynamic RT instabilities based gas pressure--density gradient inversion criterion. Purple shading marks the RS compression, identified by increasing gas pressure and decreasing radial velocity with radius.}}
    \label{fig:9}
\end{figure}

\wy{\Figure{9} presents the one-dimensional, angle-averaged radial profiles at peak luminosity, corresponding to the green curves in \Figure{5}. Radiation escaping from the shocked flow forms an RP that heats and accelerates the upstream gas, thereby reducing the velocity gradient across the SF. The associated redistribution of radiation pressure can also contribute to the deceleration of the shocked material. As the faster-moving inner ejecta encounter this decelerated flow, a reverse shock can develop or strengthen, producing an inner compression region characterized by increasing gas pressure and decreasing radial velocity. Regions that satisfy the RT instability criterion, $(dP_{\mathrm{gas}}/dr)(d\rho/dr) < 0$, indicate an unstable pressure and density stratification that can promote multidimensional mixing. Furthermore, the radiation energy density profiles in \Figure{9} indicate that radiation pressure exceeds gas pressure by approximately two orders of magnitude in all four models, highlighting the dominant role of radiation in shaping the shock structure and hydrodynamic evolution.}

Multigroup LCs help disentangle the effects of \CCSM\ geometry and mass. In model \mnd, the soft X-ray emission develops a broader and smoother shoulder because the equatorial annulus breaks the otherwise spherical \CCSM\ geometry and channels photon diffusion preferentially toward higher latitudes.

\wy{In the less dense \CCSM\ models, the soft X-ray emission peaks later than the bolometric LC. This delay arises because the radiation precursor first produces an early UV maximum together with an initial soft X-ray component ahead of the SF, whereas the main soft X-ray peak follows as radiation from the evolving, hotter shocked material diffuses through the \CCSM. The distinct temporal evolution across radiation groups arises from shock--\CCSM\ interaction, which reshapes the spectral evolution of the SBO emission.}

\subsection{Observational Prediction}
\label{subsec:dis_obs}

Direct comparison between multi-D SBO simulations and multi-wavelength observations requires energy-dependent radiation transport. Grey schemes capture the total radiative energy but cannot reproduce frequency-dependent diffusion, color-temperature evolution, or synthetic photometric LCs. Our 16-group MGFLD implementation in \CASTRO\ overcomes these limitations by resolving the emergent radiation energetics ranging from IR to X-ray.

\subsubsection{Post-Peak Cooling Curves}
In our moderately dense \CCSM\ models, \mna\ and \mnb, the breakout luminosity peaks at $t\approx60$--$75$~hr, reflecting the combined effects of shock propagation through the stellar envelope and radiative diffusion across the \CCSM. This timescale, $t\gtrsim2.5$--$3.1$~days, places the breakout emission within the early-time window accessible to wide-field transient surveys.

The flux of each band declines approximately as a power law, $F_\nu\propto t^\alpha$, where the index $\alpha$ depends on the \CCSM\ density structure after passing its maximum. The corresponding magnitude decline rate is
\begin{equation}
\frac{\mathrm{d}m}{\mathrm{d}t}
=-\frac{2.5}{\ln 10}\frac{\alpha}{t}
\approx-1.086\,\frac{\alpha}{t} \quad \mathrm{mag\ d^{-1}}\mathrm{.}
\end{equation}
For model \mna, the immediate post-peak decay indices in the NUV, bolometric, and soft X-ray bands are $\alpha \approx -1.22$, $-10.2$, and $-30$ at $t \approx 3.1$~days, respectively. As the post-shock gas cools, the soft X-ray flux declines rapidly, whereas the NUV emission is sustained by the cooling envelope.
Assuming this power-law rate holds to day~5, the magnitude decline rate is then $\mathrm{d}m/\mathrm{d}t \approx 0.26\ \mathrm{mag\ d^{-1}}$, which agrees closely with the post-maximum decline of $0.21$--$0.27\ \mathrm{mag\ d^{-1}}$ measured across the \textit{Swift}/UVOT \texttt{uvw1}, \texttt{uvm2}, and \texttt{uvw2} bands for SN~2016X \citep{2018Huang}. In contrast, model \mnb\ has a more compact \CCSM\ shell and therefore exhibits a faster post-peak decline, with $\mathrm{d}m/\mathrm{d}t\approx0.63$--$1.0\ \mathrm{mag\ d^{-1}}$ at $t=5$~days. This rapid fading is consistent with SBO models involving compact \CCSM\ extending to ${\approx}1.3\times10^{14}$~cm \citep{2016Dhungana, 2018Morozova}.

Although the shock-heated gas at late times expands homologously and transitions into the universal spherical shock-cooling decay tail \citep{2011Chevalier, 2013Sapir}, the early post-breakout emission is dominated by the dense \CCSM. This moderately dense and extended \CCSM\ in model \mna\ effectively mimics an extended stellar envelope and sustains the post-breakout UV emission. The \CCSM\ structure therefore directly influences the early LC decline and introduces degeneracy into progenitor-radius estimates based on standard shock-cooling relations, such as $L\propto R_\star E_{\mathrm exp}^{0.96}t^\alpha$ \citep{2017Waxman}. If the contribution from \CCSM\ is neglected, the enlarged effective photospheric radius may be misinterpreted as a larger stellar radius, leading to a systematic overestimate of $R_\star$ \citep{2017Dessart, 2018Foerster, 2024Irani}.
 
\subsubsection{Constraining Dense \CCSM\ Interaction: The Case of SN 2023ixf}

When a SN is embedded in a massive circumstellar envelope, as in model \mnc, the large optical depth substantially reshapes the emergent LC. The $t_{\mathrm {rise}}$ is postponed around 10 hr, and the rising duration is extended around 2-3 times, significantly longer than in the less dense \CCSM\ models.

In the NUV band, model \mnc\ reaches a peak luminosity of $3.3\times10^{41}\ \mathrm{erg\ s^{-1}}$ with a LC duration of ${\approx}60$~hr, reproducing the prolonged early UV rise observed in SN~2023ixf \citep{Hiramatsu_2023, 2023Bostroem, 2024Jacobson-Gal}.

Our MGFLD calculations show that the dense \CCSM\ model \mnc\ sustains an EUV luminosity above the helium-ionization threshold, $E\geq54.4\ \mathrm{eV}$, until $t\approx100$~hr after explosion. The ionizing emission reaches a peak luminosity of ${\approx}(2$--$3.5)\times10^{43}\ \mathrm{erg\ s^{-1}}$, whereas the EUV luminosity declines rapidly after maximum in the lower-density models. Sustained shock--CSM interaction in the dense model therefore provides a long-lived source of ionizing radiation capable of powering the persistent flash-ionization features observed in early SN spectra.

Recent spectropolarimetric observations of SN~2023ixf favor highly asymmetric and confined pre-explosion mass loss, with an inferred equatorial CSM mass of ${\approx}2\times10^{-3}\ \Msun$ \citep{2026Nagao, 2026Vasylyev}. These constraints suggest that the \CCSM\ may be strongly clumped or concentrated in a disk-like structure, similar to model \mnd. While the dense \CCSM\ in model \mnc\ absorbs high-energy photons and suppresses the resulting soft X-ray emission, the asymmetric \CCSM\ geometry of model \mnd\ allows photons to escape more efficiently through lower-density regions, producing a stronger soft X-ray LC. A hybrid configuration combining the denser \CCSM\ of model \mnc\ with the asymmetric geometry of model \mnd\ may therefore provide a more complete explanation of the multi-wavelength observations of SN~2023ixf.

\subsubsection{Shock Breakout Observability, Flash Ionization, and Light-Curve Diagnostics}
\label{subsubsec:dis_obs_xray}

Our simulations predict observational signatures of SN SBO for current and upcoming time-domain missions, including the Einstein Probe (EP) WXT ($0.5$--$4$~keV), FXT ($0.3$--$10$~keV), and the ULTRASAT NUV band ($4.3$--$5.4\,\mathrm{eV}$)~\citep{2018SYuan,2022Zhang,2024Shvartzvald}.

\begin{figure}
    \centering
    \includegraphics[width=1.0\linewidth]{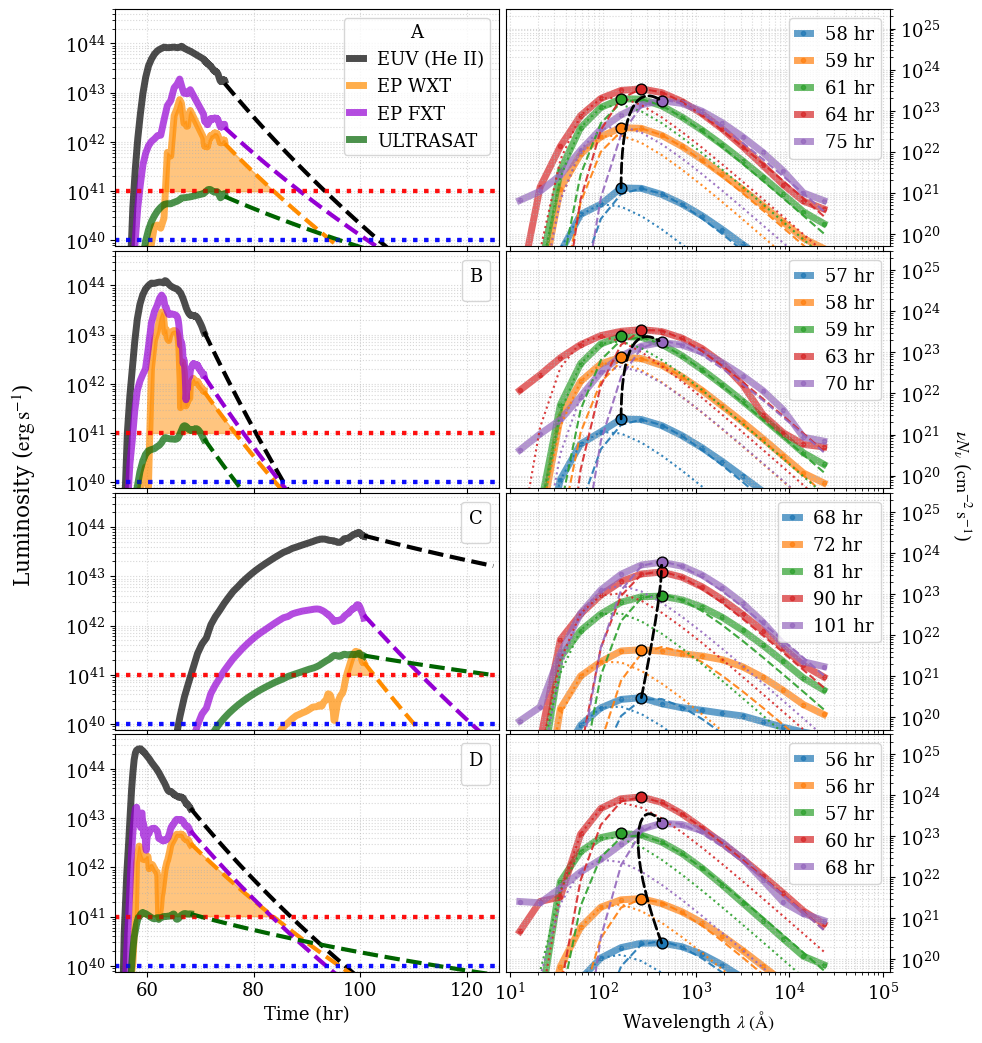}
    \caption{Left: Synthetic light curves for EP~WXT/FXT and ULTRASAT, together with the \ion{He}{2}-ionizing luminosity, calculated at a viewing angle of $\theta=90^\circ$. Dashed curves show power-law extrapolations during the cooling phase of the shocked envelope. The red and blue dotted lines mark $L=10^{41}$ and $10^{40}\ \mathrm{erg\ s^{-1}}$, respectively. The $10^{41}\ \mathrm{erg\ s^{-1}}$ threshold corresponds approximately to the luminosity detectable by EP~WXT for SN~2023ixf at $D\sim6.7$~Mpc, with the orange shaded region indicating the corresponding detection window. Right: Evolution of the SED for all models. The SED peaks shift rapidly toward the EUV during the breakout rise and subsequently redden as the ejecta cool.}
    \label{fig:10}
\end{figure}

Because our radiation groups do not fully sample energies above $1$~keV, we reconstruct the high-energy spectra using instantaneous blackbody fits to the emergent EUV and soft X-ray emission, as shown in \Figure{10}. \wy{The thermal bremsstrahlung and Comptonization in hot post-shock gas, together with possible additional plasma heating associated with a transition from a radiation-mediated shock to a collisionless shock \citep{2025Wasserman}, may produce prominent high-energy photons, causing Planckian fits to underestimate the flux above ${\approx}1$~keV, rendering our derived X-ray band luminosities and maximum detection horizons as approximate, model-dependent estimates.}\footnote{The ULTRASAT band and the photon energies required for \ion{He}{2} ionization are directly covered by our multigroup radiation grid; therefore, no blackbody extrapolation is applied to these luminosities.}

Additionally, we evaluate instantaneous line-of-sight spectra rather than LTT-convolved profiles to directly capture the intrinsic peak hardness and thermodynamic state of the breakout layer without temporal dilution.

\wy{We further extrapolate the late-time spectra to 5 days using power-law fits over the last simulated light-crossing timescale, approximately $1.6$--$2.0\,\mathrm{hr}$. Subsequent cooling of the shocked envelope may sustain NUV emission beyond the extrapolated breakout component and prolong its detectability with ULTRASAT.}

Early spectra of flash-ionized Type II SNe show \ion{He}{2} $\lambda4686\ \text{\AA}$ luminosities of ${\approx}10^{38}$--$10^{39}\ \mathrm{erg\ s^{-1}}$ sustained to $4$--$5$ days post-explosion \citep{2021Bruch, 2023Bruch}. Because the electron recombination timescale of \ion{He}{2} in the \CCSM\ is only a few minutes, maintaining this line emission requires a continuous ionizing continuum at photon energy ${\gtrsim}54.4$~eV.

To estimate the \ion{He}{2} $\lambda4686\ \text{\AA}$ recombination-line luminosity, we integrate the ionizing photon rate over $54.4$--$250$~eV. The upper limit accounts for the rapid decline of the photoionization cross section, $\sigma\propto E^{-3}$. Assuming Case~B recombination, a $4\rightarrow3$ transition probability of $0.18$, and a line-photon energy of $2.65$~eV, the approximate conversion efficiency from ionizing radiation to \ion{He}{2} $\lambda4686\ \text{\AA}$ emission is
\begin{equation}
\eta\approx\frac{2.65\ \mathrm{eV}}{54.4\ \mathrm{eV}}\times0.18\approx1\%\mathrm{.}
\end{equation}
An observed line luminosity of $L_{4686}\sim10^{38}\ \mathrm{erg\ s^{-1}}$ therefore requires a sustained ionizing luminosity of approximately $L_{\mathrm{EUV}}\sim10^{40}\ \mathrm{erg\ s^{-1}}$.

As shown in \Figure{10}, the EUV luminosities of \mna\ and \mnb\ fall below $L=10^{40}\ \mathrm{erg\ s}^{-1}$ within ${\approx}3$--$4$~days. Model \mnc\ initially produces weaker emission due to its denser \CCSM, but sustained shock--\CCSM\ interaction powers luminosities above $10^{40}\ \mathrm{erg\ s}^{-1}$ for approximately $12$--$15$~days, \wy{assuming that the power-law extrapolation continues to later times.} In the asymmetric model \mnd, the dense torus maintains the EUV luminosity above $10^{40}\ \mathrm{erg\ s}^{-1}$ for about $4$~days at the viewing angle of $\theta=90^\circ$. The lower column density along the viewing angle of $\theta=0^{\circ}$ facilitates the breakout emission and produces a prompt peak followed by a faster decay than in models \mna\ and \mnb. Assuming full ionization, the total \CCSM\ masses listed in Table~\ref{tab:name} yield maximum \ion{He}{2}~$\lambda$4686\AA\ line luminosities spanning ${\approx}10^{38}$--$10^{40}\ \mathrm{erg\ s}^{-1}$. Models \mna, \mnb, and \mnd\ are therefore consistent with the flash-ionized transients observed in \citet{2023Bruch}, whereas model \mnc\ may explain the prolonged \ion{He}{2} emission observed in shock-ionized transients \citep[$>14$~days;][]{2023Bruch}, including SN~2023ixf \citep[$7$--$8$~days;][]{2023Bostroem, 2024Jacobson-Gal}.

For targeted observations, EP~FXT reaches $F_{\mathrm{X}} \approx 10^{-13}\ \mathrm{erg\ s^{-1}\ cm^{-2}}$ in a $1\ \mathrm{ks}$ exposure and $1.0 \times 10^{-14}\ \mathrm{erg\ s^{-1}\ cm^{-2}}$ in a $10\ \mathrm{ks}$ deep follow-up, while wide-field EP~WXT monitoring reaches $F_{\mathrm{X}}\approx2.7\times10^{-11}\ \mathrm{erg\ s^{-1}\ cm^{-2}}$ and ULTRASAT reaches $F_{\mathrm{UV}} \approx 1.1\times10^{-14}\ \mathrm{erg\ s^{-1}\ cm^{-2}}$ in a $900\ \mathrm{s}$ stacked integration (Table~\ref{tab:sbo_horizons}). For a nearby event such as SN~2023ixf at ${\approx}6.7$~Mpc, the detection limits for these facilities correspond to X-ray luminosities of $L_{\mathrm{X}}\approx5.4\times10^{38}$--$1.4\times10^{41}\ \mathrm{erg\ s^{-1}}$ and a NUV luminosity of $L_{\mathrm{UV}}\approx5.9\times10^{37}\ \mathrm{erg\ s^{-1}}$.

Based on \Figure{10}, models \mna, \mnb, and \mnd\ remain detectable with WXT until approximately $t\approx3$--$3.5$~days, while FXT could extend the detection window to ${\approx}4$--$6$~days. In model \mnc, WXT detectability is delayed until $t\approx4$--$4.5$~days and is confined to a narrow window of approximately $6$~hr. Its late-time tail, however, remains detectable with FXT and ULTRASAT until $t\approx5$--$7$~days.

We estimate the detection horizon of \CCSM-mediated SBOs for each facility using the band-integrated luminosity averaged over the LC duration. The resulting horizon distances and event-rate estimates are summarized in Table~\ref{tab:sbo_horizons}. The expected detection rate is estimated as
\begin{equation}
N_{\mathrm {sbo}} \approx V_{\mathrm {sn}}\times R_{\mathrm {sn}} \times O_{\mathrm t}
\quad \mathrm{yr^{-1}},
\label{eq:detection_rate}
\end{equation}
where $V_{\mathrm {sn}}$ is the volume enclosed by the detection horizon, $R_{\mathrm {sn}}$ is the volumetric rate of core-collapse supernovae with dense \CCSM, and $O_{\mathrm t}$ is the effective sky-coverage fraction determined by the telescope field of view (FOV), $\Omega$, and survey cadence. We adopt $R_{\mathrm sn}\approx3\times10^{-5}\ \mathrm{Mpc^{-3}\ yr^{-1}}$, based on a local CCSN rate of ${\approx}10^{-4}\ \mathrm{Mpc^{-3}\ yr^{-1}}$ \citep{2011Li} and a dense-\CCSM\ fraction of ${\approx}30\%$ \citep{2021Bruch, 2023Bruch}.

For EP~WXT, our models yield detection horizons of ${\approx}8$--$84$~Mpc. Although its instantaneous FOV of $\Omega\approx3600\ \mathrm{deg^2}$ corresponds to only ${\approx}8.7\%$ sky coverage, the detectable soft X-ray emission across all our models lasts longer than EP's ${\approx}4.5$~hr survey cadence. Regular scanning covers roughly half the sky within this cadence, so we adopt an effective coverage of $O_{\mathrm{t}}\approx50\%$, yielding an expected discovery rate of $N_{\mathrm{sbo}}\approx0.04$--$37\ \mathrm{yr^{-1}}$. The broad range of $N_{\rm sbo}$ reflects the substantial theoretical uncertainties in EP detection forecasts \citep{2025Yuan}, while demonstrating the rate enhancement via a dense \CCSM.

EP~FXT, by contrast, can reach Gpc-scale horizons for targeted follow-up observations, though this estimate neglects interstellar X-ray absorption \citep{2000Wilms, 2013Willingale}. Because its narrow FOV of only ${\approx}0.8\ \mathrm{deg^2}$ makes untargeted discoveries highly unlikely, we do not estimate a serendipitous detection rate for FXT.

For the forthcoming ULTRASAT mission, the nominal detection horizon extends to ${\approx}270$--$490$~Mpc. Adopting $O_{\mathrm t}\approx0.49\%$ for its ${\sim}204\ \mathrm{deg^2}$ FOV centered on a designated sky region gives an expected rate of $ N_{\mathrm {sbo}}{\approx}12$--$74\ \mathrm{yr^{-1}}$. The lower end of this range agrees with the nominal mission forecast \citep{2014Sagiv}, while the upper end highlights the boosted UV detectability provided by dense \CCSM\ interaction.

These estimates should be regarded as upper limits, as realistic detection efficiencies are reduced by orbital visibility and other operational constraints \citep{2025Wei}.

Beyond its detectability, the X-ray LC encodes the density structure and geometry of the \CCSM. Models \mna, \mnb, and \mnd\ display the pre-maximum shoulder in \Figure{10}, providing a potential signature of \CCSM. The rising duration $t_{\mathrm {peak}}-t_{\mathrm {onset}}$ and peak epoch $t_{\mathrm {peak}}$ also vary with the \CCSM\ properties: the X-ray emission peaks at ${\approx}62$--$65$~hr in models \mna\ and \mnb, but is delayed to ${\approx}95\text{--}100$~hr in the denser model \mnc. In contrast, model \mnd\ reaches its initial peak at ${\approx}58$~hr because of its relatively low average \CCSM\ density. Subsequent interaction between the shock and the dense torus \CCSM\ produces an additional channel of X-ray emission, generating a secondary peak near $70$~hr and sustaining the late-time LC.

Finally, we identify three key signatures of SBO in a dense circumstellar environment: a distinct pre-maximum shoulder in the X-ray LC, an hour-scale delay of the soft X-ray peak relative to the EUV band, and a sustained high-energy tail of $L_{\mathrm{X}}\gtrsim10^{41}\ \mathrm{erg\ s^{-1}}$ until ${\sim}4$--$5$~days post-explosion.

\begin{deluxetable*}{lccccc}
\tabletypesize{\scriptsize}
\tablewidth{0pt}
\tablecaption{Observational Predictions\label{tab:sbo_horizons}}
\tablehead{
\colhead{Facility} & \colhead{Quantity} & \colhead{\mna} & \colhead{\mnb} & \colhead{\mnc} & \colhead{\mnd}
}
\startdata
\textbf{EP WXT} ($0.5$--$4\,\mathrm{keV}$)& $\bar{L}_{\mathrm{X}}$ & $5.43 \times 10^{42}$ & $2.27 \times 10^{43}$ & $2.27 \times 10^{41}$ & $3.67 \times 10^{42}$ \\
$F_{\mathrm{X}} = 2.7 \times 10^{-11},\ t_{\mathrm{exp}} = 1\,\mathrm{ks}$  & $d_{\mathrm{max}}$& $41.0$& $83.9$& $8.4$& $33.7$\\
$\Omega \approx 3600\,\mathrm{deg}^2,\ O_{\mathrm{t}}\approx50\%$\tablenotemark{$\ast$} & $N_{\mathrm{sbo}}$ & $4.35$& $37.1$& $0.036$& $2.41$\\
\hline
\textbf{EP FXT} ($0.3$--$10\,\mathrm{keV}$)& $\bar{L}_{\mathrm{X}}$ & $1.23 \times 10^{43}$ & $4.69 \times 10^{43}$ & $1.89 \times 10^{42}$ & $1.34 \times 10^{43}$ \\
$F_{\mathrm{X}} = 1.0 \times 10^{-14},\ t_{\mathrm{exp}} = 10\,\mathrm{ks}$ & $d_{\mathrm{max}}$& $3209$& $6263$& $1257$& $3348$\\
$\Omega \approx 0.8\,\mathrm{deg}^2$& $N_{\mathrm{sbo}}$ & \nodata\tablenotemark{$\dagger$} & \nodata\tablenotemark{$\dagger$} & \nodata\tablenotemark{$\dagger$} & \nodata\tablenotemark{$\dagger$} \\
\hline
\textbf{ULTRASAT} ($4.3$--$5.4\,\mathrm{eV}$)& $\bar{L}_{\mathrm{UV}}$ & $3.19 \times 10^{41}$ & $9.93 \times 10^{40}$ & $2.00 \times 10^{41}$ & $9.29 \times 10^{40}$ \\
$F_{\mathrm{UV}} = 1.1 \times 10^{-14},\ t_{\mathrm{exp}} = 900\,\mathrm{s}$   & $d_{\mathrm{max}}$& $492$& $275$& $390$& $266$\\
$\Omega \approx 204\,\mathrm{deg}^2,\ O_{\mathrm{t}}\approx0.49\%$& $N_{\mathrm{sbo}}$ & $74$& $13$& $37$& $12$\\
\enddata
\tablecomments{Flux limits, $F$, are given in $\mathrm{erg\,s^{-1}\,cm^{-2}}$; sustained luminosities, $\bar{L}$, in $\mathrm{erg\,s^{-1}}$; detection horizons, $d_{\mathrm{max}}$, in $\mathrm{Mpc}$; and untargeted discovery rates, $N_{\mathrm{sbo}}$, in $\mathrm{yr^{-1}}$. Here, $\bar{L}$ denotes the band-integrated luminosity averaged over the LC duration. $\Omega$ is the instantaneous FOV, and $O_{\mathrm{t}}$ is the effective fractional sky coverage.}
\tablenotetext{\ast}{The instantaneous FOV corresponds to ${\approx}8.7\%$ sky coverage. Because the detectable emission across all models outlasts EP's ${\approx}4.5$~hr survey cadence, an effective coverage of $O_{\mathrm{t}} \approx 50\%$ is adopted.}
\tablenotetext{\dagger}{EP~FXT is dedicated to pointed follow-up; hence, untargeted event rates are not applicable.}
\end{deluxetable*}
\subsection{Model Caveats and Future Perspective}
\label{subsec:dis_caveats}

Although our 2D MGFLD simulations capture multi-D hydrodynamic instabilities and multi-wavelength color evolution, several theoretical limitations remain.

First, we adopt a constant electron-scattering opacity. Realistic bound-free and line opacities derived from detailed atomic calculations, such as those performed with \code{Cloudy} \citep{2023Gunasekera}, vary strongly with frequency, density, and temperature. Including these processes could increase the optical depth of the stellar envelope and lengthen the photon-diffusion timescale. However, implementing frequency-dependent multigroup opacities is numerically challenging because of their nonlinear coupling to the gas and radiation fields. Near the SF, steep gradients in temperature, density, and radiation energy can produce large accumulated errors during implicit matrix iterations. These numerical uncertainties may outweigh the gain in physical realism relative to a stable, well-controlled, and energy-conserving treatment. \wy{A possible transition from a radiation-mediated shock to a collisionless shock is not explicitly modeled in our simulations, as the characteristic scales of energy dissipation and particle heating are far below our numerical resolution \citep{2025Wasserman}. Such a transition could introduce additional heating and radiative reprocessing, potentially modifying the emergent SED. The strength and radiative consequences of a collisionless shock depend sensitively on the residual velocity jump across the shock, which can be substantially modified by radiative pre-acceleration of the upstream gas and multidimensional mixing, as demonstrated in our 2D simulations.} 

Second, we approximate the complex \CCSM\ as \wy{prescribed initial density profiles}. In reality, several pre-explosion mechanisms can produce highly diverse circumstellar structures: interior instabilities and envelope pulsations may generate discrete, non-monotonic density enhancements \citep{2025Bronner, 2026Laplace}; wave-driven superwinds can form continuous, dense outflows \citep{2013Ofek,2026Sengupta}; and binary interactions can create strongly asymmetric geometries \citep{2015Milisavljevic,2025Matsuoka, 2026Tsai} as well as radial inhomogeneity \citep{2023Maeda,2026Maeda}. Although the detailed structure and kinematics of \CCSM\ remain uncertain, our adopted dense profiles are constrained by observations and therefore remain physically motivated \citep{2017Yaron}.

Third, turbulent mixing is intrinsically three-dimensional due to the different energy cascades in 2D and 3D flows. Nevertheless, 2D simulations remain effective for resolving the dominant mixing processes at substantially lower computational cost. Previous SN calculations have shown that 2D models can reproduce RT mixing depths and large-scale gas dynamics broadly comparable to those obtained in 3D \citep{2010Joggerst, Ken2023ApJ}. Our simulations therefore capture the principal mixing-driven effects on photon transport and produce LCs consistent with published 3D calculations of SBO from RSGs \citep{2022Goldberg}.

\section{Conclusions}
\label{sec:conclusions}

\wy{We have performed the first two-dimensional multigroup radiation-hydrodynamic simulations of shock breakout from red supergiants embedded in confined circumstellar media (\CCSM), using the MGFLD framework in \CASTRO. Our results demonstrate that the coupled evolution of the forward shock, radiating gas shell, and optically thick \CCSM\ governs the propagation and escape of breakout radiation, shaping its temporal and spectral evolution.} 

The color evolution of emission undergoes rapid blue-shift during the breakout rise, with the emission peaking in the EUV and soft X-ray, followed by gradual reddening during post-breakout cooling. Secondary soft X-ray features trace the emergence of hotter, deeper shock-heated material through the \CCSM. 
Across the model grid, the bolometric peak luminosities span ${\approx}1$--$3\times10^{44}\ \mathrm{erg\ s^{-1}}$, with LC durations of ${\approx}4$--$35$~hr. Increasing the density or radial extent of the \CCSM\ systematically delays, suppresses, and broadens the breakout emission, producing empirical correlations between the observable light-curve properties and the circumstellar geometry.

Our results demonstrate that multi-dimensional, wavelength-dependent radiation-hydrodynamic modeling is essential for interpreting early supernova emission. By connecting multigroup SBO light curves with constraints from flash spectroscopy, this study provides a quantitative framework for high-cadence UV and X-ray surveys, including Swift, ULTRASAT, and Einstein Probe. Future simulations incorporating fully three-dimensional dynamics and detailed atomic opacities will further improve constraints on the mass-loss environments of massive stars immediately before core collapse.

\begin{acknowledgments}
We thank Sung-Han Tsai for useful discussions and Masaomi Ono and Po-Sheng Ou for providing the presupernova model. This research is supported by the National Science and Technology Council, Taiwan, under grant No. NSTC 113-2112-M-001-028-, 114-2811-M-001-094, NSTC 115-2112-M-001-001- and the Academia Sinica, Taiwan, under a career development award under grant No. AS-CDA-111-M04. KC acknowledges the support of the Alexander von Humboldt Foundation. WC acknowledges support of the National Taiwan University Scholarship for Direct Pursuit of Doctoral Degree, the Scholarship of the Chung Hwa Rotary Education Foundation (Rotary International District 3490, Sanchung–Sanyang Rotary Club), 2025 National Center for Theoretical Sciences Student Outstanding Paper Award, and the 2026 National Science and Technology Council--Deutscher Akademischer Austauschdienst (NSTC--DAAD)--Sandwich--Scholarship Programme. KM acknowledges support from the Japan Society for the Promotion of Science (JSPS) KAKENHI grant Nos. 24KK0070 and 23H04894. 
Our computing resources were supported by the National Energy Research Scientific Computing Center (NERSC), a U.S. Department of Energy Office of Science User Facility operated under Contract No. DE-AC02-05CH11231, and the KAWAS Cluster at the Academia Sinica Institute of Astronomy and Astrophysics (ASIAA). The work of FKR is supported by the Klaus Tschira Foundation, by the Deutsche Forschungsgemeinschaft (DFG, German Research Foundation) -- RO 3676/7-1, project number 537700965, and by the European Union (ERC, ExCEED, project number 101096243). Views and opinions expressed are, however, those of the authors only and do not necessarily reflect those of the European Union or the European Research Council Executive Agency. Neither the European Union nor the granting authority can be held responsible for them. 
\end{acknowledgments}
\bibliography{reference.bib}
\bibliographystyle{aasjournalv7}
\end{CJK*}
\end{document}